\documentstyle[aaspp4]{article}

\begin{document}

\title{Surface Photometric Calibration of the Infrared Tully-Fisher
	   Relation Using Cepheid-based Distances of Galaxies}

\author{Masaru Watanabe\altaffilmark{1}}
\affil{Japan Science and Technology Corporation, Tokyo 102-0081, Japan}
\authoremail{watanabe@aquarius.plain.isas.ac.jp}

\author{Naoki Yasuda}
\affil{National Astronomical Observatory of Japan, Tokyo 181-8588,
       Japan}
\authoremail{naoki.yasuda@nao.ac.jp}

\author{Nobunari Itoh}
\affil{Kiso Observatory, University of Tokyo, Nagano 397-0101, Japan}
\authoremail{nitoh@kiso.ioa.s.u-tokyo.ac.jp}

\altaffiltext{1}{Postal address: Center for planning and information
		 systems, Institute of Space and Astronautical Science,
		 Kanagawa 229-8510, Japan}\hspace{5mm}

\begin{center} {(\em Submitted to Ap.J.}\/)\end{center}

\begin{abstract}
Infrared $J$ and $H$ surface photometry are carried out for nearby 12
galaxies whose distances have been accurately measured via HST Cepheid
observations. Using the total, isophotal, and surface-photometric
aperture magnitudes we calibrate the infrared luminosity-line width
relation (IRTF). It is found that IRTF changes its slope
at $\log W_{20}^c\sim 2.45$ in all the examined magnitude systems. The
apparent scatter of IRTF is not significantly reduced when surface
photometric magnitudes are used instead of the conventionally used
synthetic aperture magnitude $H_{-0.5}$. It is also shown that the color
$(I-H)_T$ of the nearby calibrator galaxies is redder by $\sim 0.2$\,mag
than the Coma cluster galaxies, but such a trend is not clearly visible
for the Ursa Major mostly because of poor statistics. The color offset 
of the Coma is analogous to that previously found in $I_T-H_{-0.5}$. 
From the present calibration of $H$-band IRTF, 
we obtain the distance to the Coma cluster to be 
$m-M=34.94\pm 0.13$\,mag, where no account is taken of the $I-H$ color 
problem. Using the CMB-rest recession velocity of the Coma cluster we 
obtain $H_0=73\pm 4$\,km\,s$^{-1}$\,Mpc$^{-1}$.

\end{abstract}


\section{Introduction}

The infrared luminosity-line width relation (hereafter IRTF) has been
studied conventionally using the $H$-band synthetic aperture magnitude
denoted as $H_{-0.5}$ (Aaronson et al 1986, Pierce \& Tully 1988,
Freedman 1990, Sakai et al 2000). Originally introduced by Aaronson,
Mould \& Huchra (1979), the $H_{-0.5}$ magnitude was defined as the flux
observed within an aperture having a diameter $A=10^{-0.5}D_{25}$, where
$D_{25}$ stands for the optical isophotal diameter measured at the
surface brightness level of $\mu_B=25$\,mag\,arcsec$^{-2}$. Based on
this rather heterogeneous magnitude system, the $H$-band IRTF has been a
peculiar one compared with the optical TFs which are all based on the
ordinary surface photometric magnitudes (e.g., Giovanelli et al. 1997b).

The $H_{-0.5}$ system was originally contrived from an observational
limitation that a 2-D imager was not available in the infrared
wavelength domain. Even after the advent of an infrared 2-D imager, its
physical size had been insufficient for
accurate surface photometry. In the meanwhile homogeneous data sets
of $H_{-0.5}$ had been accumulated for both field and
cluster galaxies (Aaronson et al. 1982, Bothun et al. 1985, Aaronson et
al. 1989, Tormen \& Burstein 1995 [hereafter TB95]). Because of their 
usefulness these
data have played a pivotal role in the subsequent IRTF studies, and
consequently led to the present paradigm of $H_{-0.5}$ system for IRTF
studies. Nevertheless, it is also true that we still await IRTF to be
calibrated with a surface photometric magnitude, as it provides a more
fair and thus preferable measure of a galaxy luminosity.
This requirement will be imminent and inflated as a significant progress
will be made in
extensive infrared surveys such as the 2-Micron All-Sky Survey (2MASS,
Skrutskie et al. 1997) and the Deep Near-Infrared Southern Sky (DENIS)
survey (Epchtein et al. 1994).

The infrared surface
photometry of these galaxies is important not only for the calibration of
the IRTF, but also for exploring the photometric properties of the nearby
galaxies.
Recently Sakai
et al (2000; hereafter S00) has raised a question pertaining to the
possible offset of the $H$ magnitude between the nearby and cluster
galaxies. In that paper they calibrated the luminosity-line width
relations in $BVRIH_{-0.5}$ and applied them to 23 clusters to obtain
the global value of $H_0$. From an internal comparison between $H_0$'s
obtained in these different wavelengths they found that the mean $H_0$
obtained in $H_{-0.5}$ is 10\% smaller than those obtained in the
others. Examining several observational properties of the sample
galaxies they came to a conclusion that there should be a systematic
difference in $H_{-0.5}$ between the nearby calibrators and the cluster
galaxies. It is yet unclear, however, whether this difference somehow
comes from this peculiar photometry technique of $H_{-0.5}$ or from a
difference of intrinsic photometric properties between these galaxies.
In order to disentangle this problem it is desirable to obtain surface
photometry data for these nearby galaxies.

To address these issues we carry out $J$ and $H$ surface photometry
for nearby galaxies having HST Cepheid distance measurements. These
galaxies have a large apparent size of $3-10\,$arcmin in diameter.
Accurate imaging photometry of such large galaxies is made available
with a large-format infrared camera attached to the Kiso Schmidt
telescope (Itoh, Yanagisawa \& Ichikawa 1995). With this configuration
the camera covers an 18 arcmin-square field of view. Using the data we
calibrate IRTF and apply the relation to cluster galaxies. Some $H$-band
photometric properties of galaxies will be also examined including the
$H$ magnitude problem mentioned above.

Contents of this paper is as follows. Observations of the sample
galaxies and literature data used are described in section
\ref{sec:sample}. Photometry results, accuracy, and external consistency
are presented in section \ref{sec:data}. Calibration of IRTFs is given
in section \ref{sec:calibration}, followed by a discussion on
photometric properties of the sample galaxies and the application of
IRTF to cluster galaxies given in the literature in section
\ref{sec:discussion}.

\section{The Sample and Observations}
\label{sec:sample}

\subsection{The Sample}

Our sample comprises nearby 12 spirals (Table \ref{tbl-1}). They have
been selected according to two criteria that HST Cepheid observations
be available and that inclination be larger than $45^\circ$ according to
S00 or RC3. The literature data of these galaxies are also summarized in
Table \ref{tbl-1}. In order to be as consistent with S00 as possible, we
refer to the compilation by S00 for all the external data including
galaxy distances, H {\sc i} line widths, and inclinations whenever available.

As for the extinction correction we adopt Schlegel et al (1998) for the
Galactic extinction and Tully et al (1998) for the internal extinction.
These were both adopted by S00 as well. Since Tully et al (1998) did not
actually provide extinctions in $J$ and $H$, these values are evaluated
using relations $A_i^{(H)}=0.5A_i^{(I)}$ given by S00 and
$A_i^{(J)}=0.8A_i^{(I)}$ obtained by ourselves from an interpolation
between $I$ and $H$. S00 does not include NGC\,4639, therefore its H
{\sc i} line width data are taken from Huchtmeier \& Richter (1989), and the
inclination is computed using the axial ratio given in RC3 and the
inclination formula given in S00.

As for the extinction corrections, there have been several alternative
prescriptions advocated by different authors. To examine the effect of
adopting different prescriptions we compare our adopting $A_g$ and $A_i$
with those computed from Burstein \& Heiles (1984) and RC3,
respectively. Note in advance that 
since the uncertainty in the absolute magnitudes
of galaxies due to the distance error is already more than 0.1 mag, that
due to differences in various extinction correction methods is
negligible.
The Galactic extinction by Burstein \& Heiles (1984) is presented in $B$
hence converted to those in $J$ and $H$ using the ratio of
$A_g^{(B)}:A_g^{(J)}:A_g^{(H)}=4.315:0.902:0.576$ given by Schlegel et
al (1998). This comparison shows that the Burstein's value is
systematically different from our $A_g$ only by less than 0.02\,mag,
which is of no significance in the present study. For the comparison of
the internal extinction the $B$-band value given in RC3 is also
converted to those in $J$ and $H$. In this conversion we are based on
the ratio given by Tully et al (1998). They provide
$A_i^{(I)}/A_i^{(B)}=0.59$ which, in conjunction with the
above-mentioned relations between $A_i^{(I)}$, $A_i^{(J)}$, and
$A_i^{(H)}$, leads to a ratio of
$A_i^{(B)}:A_i^{(J)}:A_i^{(H)}=1:0.47:0.30$. This ratio is
substantially different from that previously given for $A_g$. According
to Han (1992) and Tully et al (1998), this difference is mostly ascribed
to the configurations of the Galactic and internal extinction which
compel a light ray to pass through a different path lengths toward us.
From this conversion it is shown that the systematic difference of $A_i$
between RC3 and our adopted scheme is mostly less than 0.05\,mag, except
for only NGC\,7331 which raises 0.1\,mag difference. These differences
are again negligible in the present work. As emphasized in S00, which
scheme to choose is insignificant but a consistency is important in a TF
analysis.

\subsection{Observations}
\label{sec:observations}

The $J$ and $H$ images were obtained using the 105cm Schmidt telescope
at the Kiso Observatory during the period between 1998 October and 1999
October.
Because of a time limitation a $J$ image of NGC\,4536 could not be
obtained. The telescope was equipped with a large-format near-infrared
camera called the Kiso Observatory Near Infrared Camera or KONIC as
abbreviated (Itoh, Yanagisawa \& Ichikawa 1995). KONIC is a PtSi camera
having 1040$\times$1040 physical pixels. The output data are binned into
1040$\times$520 pixels, while we use the image further binned into
520x520 pixels. The camera mounted on the Kiso Schmidt covers
18\,arcmin-square FOV hence has a pixel scale of 2.12\,arcsec par the
$2\times 2$ binned pixels.

A single on-source exposure for a targe galaxy was taken to be 300\,sec.
The same duration was assigned to an off-source exposure for a blank sky
region displaced by at least 20\,arcmin from the target galaxy. The
off-source exposure was taken after every 2 on-source exposures.
Typically successive 12 on-source exposures plus 6 intervening
off-source exposures were thus taken for a single galaxy. The number of
the on-source exposures actually varied from 4 to 16 depending on the
galaxy brightness and/or a sky condition of the night. Along with these
exposures, infrared standard stars of Elias et al (1982) were also
observed every night at various zenith distances for the magnitude
calibration. The exposures for the standard stars were appropriately
defocused so that we could accumulate a sufficient number of the
background signal count to avoid a deferred
charge problem of the camera (Itoh, Yanagisawa \& Ichikawa 1995).

\subsection{Data reduction}

The observed data were reduced using an image reduction software
IRAF\footnote{IRAF is distributed by the National Optical Astronomy
Observatories, which is operated by the Association of Universities for
Research in Astronomy, Inc. (AURA) under cooperative agreement with the
National Science Foundation.} V2.11.3 and STSDAS V2.1. A dark count was
estimated and subtracted from the raw image using the Richardson's
formula calibrated with the dark images. Image flattening was performed
in a usual manner using combined blank sky images. A reduction error in
these procedures are discussed later in section \ref{sec:accuracy}. A
sky background region in the galaxy frame was determined as an exterior
region of a circle centered at the target galaxy. The radius, which
extended up to $\sim 7$\,arcmin for the largest galaxies such as
NGC\,925, 4725, and 7331, was determined according to the apparent size
of the galaxy. Signal counts of the sky background region was fitted
with polynomial surfaces of several orders, and the best one determined
by the visual inspection into the residual counts was subtracted from
the image. Actually a first-order polynomial surface was sufficient for
most of the images. The remaining signal counts were then corrected for
the atmospheric extinction and converted into the count in the CIT
magnitude system using the standard star data reduced in a similar way.
In this conversion the extinction coefficients were determined for each
individual night separately while the color conversion coefficients were
commonly taken from the recommended color equations
$J_{Kiso}=J_{CIT}+0.06(J-H)_{CIT}$
and $H_{Kiso}=H_{CIT}+0.10(J-H)_{CIT}$.

Considering an undesirably extended psf of the camera (Yanagisawa, Itoh
\& Ichikawa 1996) we deconvolved the galaxy images using a psf image.
The deconvolution was performed using the Lucy-Richardson's method
realized in the STSDAS package. In this deconvolution two different
psf images were separately applied to the same galaxy images to see an
internal consistency. One psf is composed with several undisturbed
stellar images in a galaxy frame, while the other is taken from a
template psf image created earlier for an instrumental verification
purpose. Each of the psfs has its own advantage over the other; the
former has the same seeing condition as have the galaxy image to be
working on, while the latter has a sufficiently high S/N hence
reproduces the extended fainter outskirt of
the psf more clearly. Performing these two deconvolutions we obtained
alternative results, i.e., the former experiment reproduces the
galaxy aperture data which are in better agreement with the literature
data as used in section \ref{sec:external consistency} while the latter
provides better rectification of galaxy images without any discernible
psf outskirt even around bright galaxies. The worse agreement of the
aperture data for the latter deconvolution experiment is that the
deconvolved galaxy image provides systematically too bright a magnitude
at the inner apertures. It is noted that this systematic discrepancy,
amounting typically to $\sim 0.1$\,mag, is unlikely due to a
difference of the seeing disk conditions between our observations and
those for the aperture photometry of the literature data; this is
because the discrepancy remains up to an aperture whose diameter is a
few tens of seconds, where the difference of the seeing size should be
no more an important issue on the aperture magnitude.

In spite that these two experiments of deconvolution provide partly
unsatisfactory results, they nevertheless give consistent results
regarding global photometry parameters such as the total, isophotal,
and,
synthetic aperture magnitudes that we are mostly concerned with in this
study. Making best use of this preferable result, we constructed a
moderate psf image merging the kernel of the former psf and the extended
cross and ring-like pattern of the latter. We find that the composite
psf provides the most satisfactory deconvolution result in the sense
that it simultaneously reduces the two undesirable aspects of the
previous deconvolutions. Accordingly, the composite psf was created for
and applied to each individual galaxy for the deconvolution.

After the deconvolution foreground stellar images were cleaned out using
the interpolation from the surrounding sky background count to provide
the final galaxy image for photometry. None of the foreground stars
severely contaminates the target galaxies, thereby we may disregard any
photometry uncertainty to be introduced in this cleaning procedure. The
deconvolution often generated a ring-shaped negative overshoot around
bright stars. The amplitude of the overshoot was however negligible
compared with the gross internal photometry error, so these defects were
just cleaned in the same light of cleaning the inner stellar image.

The amplitude of the sky background shot noise is typically equivalent
to the surface brightness levels of $\mu_J = 21$\,mag\,arcsec$^{-2}$ and
$\mu_H = 20$\,mag\,arcsec$^{-2}$. In order to increase S/N of the
fainter part of the galaxies we smoothed the image using a
5\,pixel-square or 11\,arcsec-square boxcar window. On the smoothed
image we computed a series of isophotes at various surface brightness
levels and fitted ellipses to the isophotes. The luminosity profiles and
contour maps are given in Fig.\ref{fig1}. We compute several
isophotal parameters at $\mu_J=22$\,mag\,arcsec$^{-2}$ and
$\mu_H = 21$\,mag\,arcsec$^{-2}$, which are collectively presented later
in section \ref{sec:data}. A caution is needed for the map of NGC\,925
in $J$, as the isophotes is less reliable at its north-eastern
outskirts. That portion of the image is unfortunately disturbed by a
deferred charge in some pixel rows proximity to the overscan area 
attached
to the boundaries of each quadrant of the detector array. This degradation of the
image may affect the accuracy of some isophotal geometrical parameters.
However, the fraction of the disturbed light is quite small compared
with the galaxy total light, so we presume that the $J$ photometry of
NGC\,925 does not suffer any additional serious uncertainty from this
peripheral image degradation.

To obtain asymptotic total magnitudes $J_T$ and $H_T$ we extrapolated
the exponential disk to infinity. In this extrapolation the disk part
of the galaxies was determined from the visual inspection into the
luminosity profiles. To verify this determination we also referred to
the axial ratio and the position angle varying along with the isophotal
radius. The extrapolation typically amounts to 0.05\,mag with a few
exception of faint or short exposure galaxies for which we need
$0.2-0.3$\,mag of extrapolation. The extrapolation is subject to the
uncertainties of the axial ratio, disk scale length, and central
surface brightness adopted for the exponential disk. The error is
assessed later in section \ref{sec:accuracy} as a source of a
photometry error for $J_T$ and $H_T$.

\section{Surface photometry data}
\label{sec:data}

\subsection{Data presentation}
\label{sec:data presentation}

The results of the surface photometry are summarized in Table
\ref{tbl-2}.
Description of the columns are as follows;
column(1)  NGC number,
column(2)  Asymptotic total magnitude   $J_T$      and its error,
column(3)  Isophotal  magnitude         $J_{22}$   and its error,
column(4)  Isophotal  major axis length $D_{J22}$    in arcmin,
column(5)  Effective radius             $r_e^{(J)}$ in arcmin,
column(6)  Minor-to-major axial ratio   $R_{J22}$,
column(7)  Asymptotic total magnitude   $H_T$      and its error,
column(8)  Surface photometric aperture magnitude $H_{-0.5}^s$ with the
modified aperture of TB95,    and its error,
column(9)  Isophotal  magnitude         $H_{19}$   and its error,
column(10) Isophotal  magnitude         $H_{21}$   and its error,
column(11) Isophotal  major axis length $D_{H19}$    in arcmin,
column(12) Isophotal  major axis length $D_{H21}$    in arcmin,
column(13) Effective radius             $r_e^{(H)}$ in arcmin,
column(14) Minor-to-major axial ratio   $R_{H21}$.
Derivation of the quoted photometry error is discussed in the subsequent
section. The nominal disagreement between $R_{J22}$ and $R_{H21}$ is 
mostly ascribed to the measurement uncertainty; it is evaluated to be 
$\sim 0.05$ based on their scatter within the surface brightness levels
of $\mu_J=22$\,mag\,arcsec$^{-2}$ and of 
$\mu_H = 21$\,mag\,arcsec$^{-2}$.

\subsection{Photometric Accuracy}
\label{sec:accuracy}

\subsubsection{Internal photometry error}

Sources of the internal photometry error differ between the asymptotic,
isophotal, and aperture magnitudes. For the asymptotic magnitude $J_T$
and $H_T$ the error $\Delta$ is estimated as
$\Delta^2 = \Delta_{flt}^2 + \Delta_{clb}^2 + \Delta_{ext}^2$ where
$\Delta_{flt}$, $\Delta_{clb}$, and $\Delta_{ext}$ represent the errors
caused by the flattening procedure, magnitude calibration procedure, and
the disk extrapolation procedure, respectively. For the remaining
magnitudes such as $J_{22}$,
$H_{19}$, $H_{21}$, and $H_{-0.5}^s$ the internal error should be
$\Delta^2 = \Delta_{flt}^2 + \Delta_{clb}^2$.
Each of these error sources is described in what follows.

\paragraph{Flattening}

We computed a dark count of an object frame using the Richardson's
formula calibrated with the data taken from a number of dark frames. The
scatter of the data around the fitted formula is typically 10\,ADU.
A signal count of a single sky flat frame of a 300\,sec exposure was, on
the other hand, generally larger than 2000\,ADU/pixel in $J$ and
5000\,ADU/pixel in $H$ excluding the bias and dark counts. Accordingly,
the 10\,ADC error of the dark count subtraction causes flattening errors
less than 0.5\,\% and 0.2\,\% in $J$ and $H$, respectively. A more
serious error in the flattening procedure is brought by a temporal
variation of a local sky background brightness. To examine this
variation we divided the associated off-source sky background images
each other and see the distribution of the data counts in the quotient
frame. From this examination no discernible global gradient of the sky
background count is found for all the associated sky frames. A
discernible deviation from the mean count is of rather random nature.
The fractional deviation amounts to 3\,\% at most both in $J$ and $H$ at
the central part of the frames where a target galaxy was always exposed.
As the deviation is for a single object frame, we summed it up
quadratically for all the combined galaxy frames. From this, we obtain
the uncertainty to be 1.5\,\% for a combination of 4 frames (1200\,sec
exposure, the shortest in the present observations) and 0.8\,\% for a
composition of 16 frames (4800\,sec exposure, the longest).

\paragraph{Extrapolation of the exponential disk}

Extrapolation of the exponential disk is parameterized by the disk scale
length, disk central surface brightness, and the axial ratio. The former
two components are derived from a linear fit to the luminosity profile,
while the axial ratio is taken from S00. Adoption of this axial ratio is
because the optical isophote refers to a deeper surface brightness
level and because we like to avoid a possible unknown systematic error
that may be introduced in our axial ratio through the deconvolution
procedure. We computed the extrapolation error by practically changing
these parameters within a reasonable range. The error is obtained to be
$0.03-0.10$\,mag depending on the apparent brightness of the galaxies.

\paragraph{Magnitude calibration}

The magnitude calibration error is estimated as the scatter of the
standard star data around the equation of atmospheric extinction
correction and color transformation. The scatter is typically 0.03\,mag.

Taking all the errors given above into account we evaluate the internal
photometry error for individual galaxies. The value is given for each
magnitude in Table \ref{tbl-2}. We achieve a typical accuracy of
$0.06$\,mag for the total magnitude except for two faintest galaxies
NGC\,2541 and 3319. The accuracy is even better for aperture and
isophotal magnitudes as they refer only to relatively brighter portion
of the galaxy image.

\subsection{External consistency}
\subsubsection{Photometry}
\label{sec:external consistency}

Growth curves of each galaxy are computed using a series of circular
apertures. We compare the curves with aperture photometry data given
in the literature to check an external consistency. The literature data
include Glass (1976), Aaronson (1977), Peletier \& Willner (1991),
TB95, and the 2MASS extended
source catalog available on the web
(http://irsa.ipac.
caltech.edu/applications/CatScan/).

The result of the comparison is shown in Fig.\ref{fig2}. Most of the
literature data show an agreement with our values within 0.1\,mag. A
relatively large and systematic discrepancy worth mentioning is observed
in the following four galaxies;
NGC\,3198 : Aperture data of TB95 (2 apertures) are systematically
            fainter by 0.15\,mag than our photometry.
NGC\,3319 : Aperture data of TB95 (3 apertures) are systematically
            fainter by 0.1\,mag than our photometry.
NGC\,4639 : We find an increasing discrepancy toward larger apertures
			between surface photometric aperture data of 2MASS (9
			apertures) and ours. The discrepancy amounts to
			$\sim 0.3$\,mag at the outermost aperture. Aperture data of
			Peletier \& Willner (1991) (1 aperture) are more consistent
			with the 2MASS data sequence. On the other hand, the
			template curve of TB95 for this type of galaxy ($T=4$)
			matches better with our growth curve at the outer slope,
			implying that the 2MASS photometry is not sufficiently deep
			for this galaxy.
NGC\,4725 : Aperture data in $H$ by TB95 (5 apertures) are fainter by
            0.1\,mag than our photometry.
As for these four galaxies, independent data are needed to give further
discussion on this photometry result. There are two other galaxies which
show large discrepancy regarding $H_{-0.5}$;
NGC\,4535 : $H_{-0.5}$ of TB95 is fainter by 0.29\,mag than our
            $H_{-0.5}^s$. This error on $H_{-0.5}$ is ascribed to its
			wavy growth curve which cannot be well fitted with the
			monotonically increasing template curve of TB95.
NGC\,7331 : $H_{-0.5}$ of TB95 is brighter by 0.36\,mag than our
            $H_{-0.5}^s$. This error on $H_{-0.5}$ is attributed to the
			lack of aperture data having a larger diameter.

\subsubsection{Axial ratio}

We compare the axial ratio with those obtained by S00. The agreement
is better than $\Delta R=0.05$ for most of the galaxies. Relatively
large error is found for NGC\,3319 ($R_{J22}=0.37$ and $R_{H21}=0.32$,
while $R_{S00}=0.51$). The compilation of the kinematical 
inclinations by S00 indicate $R=0.53$, which is in better agreement 
with $R_{S00}$. This large discrepancy of the axial ratio of NGC\,3319
is ascribed to its faint, loosely-wound spiral arms. In the infrared 
images these arms are hardly visible, therefore they give negligible 
contribution to determining the axial ratio. This suggests that the
optical image is generally perferable for determining the inclination
of spirals. 

\section{Calibration of the infrared luminosity-line width relations}
\label{sec:calibration}

IRTF diagrams calibrated in several magnitudes are shown in
Fig.\ref{fig8}. A striking feature common to all of the diagrams is that
the IRTF changes its slope at around $\log W_{20}^c\sim 2.45$. A further
discussion on this issue is given in the subsequent section using an
enlarged sample.
Results of the linear regression fit characterized by the slope, zero
point, and the apparent scatter, are summarized in Table \ref{tbl-3}.
These parameters are computed both for the entire sample and for the
subsample limited to
$\log W_{20}^c>2.45$. Note a considerable reduction of the apparent
scatter for the subsample compared with that of the entire sample.

The calibration result with $H_{-0.5}^s$ is apparently in good agreement
in slope, zero point, and apparent scatter compared with that obtained
by S00. It should be noted that S00 referred to Aaronson et al (1982)
for the synthetic aperture photometry scheme which is different in the 
aperture system and the growth curve templates from the TB95 scheme that
we adopt for $H_{-0.5}^s$. Accordingly, the present agreement 
implies that the photometry schemes of Aaronson et al (1982) and TB95
have been both adequately accurate to obtain $H_{-0.5}$ for the purpose 
of applying it to IRTF, given the current uncertainty of other 
observables such as the line width, inclination, and extinctions. 

\section{Discussion}
\label{sec:discussion}

In this section we add Coma cluster data by Bernstein et al (1994) and
Ursa Major (UMa) cluster data by Peletier \& Willner (1993) to the
discussion in order to enlarge the sample and to apply the present
calibration result to. Both of the photometry data were calibrated
using the Elias's standard stars hence have a zero point comparable
with our data. Bernstein et al (1994) provided $H_T$ for 21
galaxies, among which 12 galaxies are also given $H_{-0.5}$ by TB95.
Galaxy inclination and H {\sc i} line width data are taken from their original
measurement and compilation. Peletier \& Willner (1993) provides $H_T$,
$H_{-0.5}^s$ and $H_{19}$ for 22 galaxies. Inclination and the line
width are taken from their paper. Corrections for the Galactic and
internal extinctions are performed following the manner we applied to
the present sample.

\subsection{Photometric properties of $H_{-0.5}$}

We first examine the difference between TB95's aperture magnitude,
$H_{-0.5}$ and our $H_{-0.5}^s$ as a function of a galaxy
inclination and of type to see any systematic errors in
$H_{-0.5}$. They both show no significant dependency, which means
that the error of $H_{-0.5}$ predominantly comes from a random
error.

We examine dependencies of $H_{-0.5}-H_T$ on a galaxy axial ratio and on
type (Fig.\ref{fig05}). Neither early nor late spirals exhibit a
significant dependency of $H_{-0.5}^s-H_T$ on the axial ratio. As is
qualitatively expected from the variety of the bulge-to-disk light ratio
on galaxy types, the mean values of $H_{-0.5}^s-H_T$ is different
between these two type ranges; $H_{-0.5}^s-H_T=0.62\pm 0.05$\,mag for
$T\le 4$ while $H_{-0.5}^s-H_T=0.77\pm 0.06$\,mag for $T\ge 5$. This
difference implies that the application of IRTF in $H_{-0.5}$ need a
caution; if the calibrators and the cluster samples consist of
considerably different population in terms of galaxy types, these may
lead to a distance estimate suffering from a systematic error.

A rather curious implication from this data plot is a deficiency of late
spirals with $H_{-0.5}^s-H_T>0.5$ at $0.3<D_{min}/D_{maj}<0.5$. We are
yet to be confident of its reality, however, and certainly think that we
need much more data for any further discussion. If this ever turns out
to be the case, the rather moderate range of the axial ratio will make
it an interesting and challenging problem.

\subsection{Environmental dependence of galaxy color}

S00 demonstrated that there is a significant discrepancy between $H_0$'s
obtained in $I$ and $H_{-0.5}$ which corresponds to a systematic
difference by 20\% of the sample cluster distances. In a close relation
to this it is also shown in S00 that the nearby calibrators have a
redder color in $I_T-H_{-0.5}$ ( but not in $(B-I)_T$ ) than the Virgo,
Ursa Major, Coma, and two other clusters for a fixed $\log W_{20}^c$.
We confirm that this $I_T-H_{-0.5}$ color offset persists even if we use
$H_{-0.5}^s$ for the nearby calibrators. Instead of $I_T-H_{-0.5}$, we
plot $(I-H)_T$ in Fig.\ref{fig06}. Although the Ursa Major galaxies show
a relatively wide distribution regardless of $\log W_{20}^c$ and the
sample seems yet poor, the plot for the Coma galaxies shows with more
convinction that the nearby calibrators are likely redder than the 
cluster galaxies. This fact indicates that some intrinsic color offset 
really exists in $I-H$ between the nearby calibrators and a certain 
cluster of galaxies. To see whether this
color offset is caused alone by the possible peculiar properties of the
$H$ luminosity, we plot another color $(B-H)_T$ in Fig.\ref{fig08}. In
this figure nearby calibrators are again found to be redder than the
Coma galaxies but not so compared with the UMa. These experiments
demonstrate that, although the color offset in $I-H$ is quite likely to
be real, we may not come to a hasty conclusion that the nearby
calibrators are always redder than cluster galaxies in the colors
relevant to $H$. To address this color problem further in detail we need
to accumulate more $H$ photometry data and examine the relevant colors
reflecting the cluster environment.

\subsection{Implication of properties of IRTF}
\label{sec:IRTF_property}

Whether the TF relation is linear for a respectable range of the line
width or not has been an issue of controversy; while most of the TF 
studies have assumed a linearity in TF relation, some studies 
(e.g., Aaronson et al 1986; Mould, Han \& Bothun 1989; McGaugh et al 
2000) adopted or argued for the nonlinearity of the relation.

As shown in Fig.\ref{fig8}, our result of IRTF calibration apparently 
exhibits a breakdown for fainter galaxies with $\log W_{20}^c<2.45$ 
regardless of the bandpass or magnitude system. This boundary may be 
comparable with a similar break at 
$V_c(\sim W_{20}^c)\sim 90$\,km\,s$^{-1}$ demonstrated by McGaugh et al 
(2000). The IRTF diagrams for the Ursa Major
clusters are shown in Fig.\ref{fig07_uma}, where the apparent magnitude
is used instead. Overplotted are the IRTFs of the nearby calibrators as
shown in Fig.\ref{fig8} but vertically scaled so that the gross scatter
be minimized at the range $\log W_{20}^c\ge 2.45$. Note in passing that 
this procedure yields simultaneously the cluster distance and the 
slope of the IRTF for the composite sample of galaxies, to which we will
be back in the next section.
It is evident from
these diagrams that the breakdown of IRTF is reproduced by the UMa
galaxies. 
  Note also that the distribution of nearby calibrators and the cluster
  galaxies for $\log W_{20}^c<2.45$ are naturally homogenized as well.
Although the sample is not definitive yet, these facts provide another 
support to that the breakdown is a common features of IRTF irrespective 
of galaxy environment. It should be noted, however, that the 
distribution of these fainter galaxies exhibits a larger scatter and a 
steeper slope which
render a separate linear fitting to these data almost meaningless
compared with the brighter galaxies with $\log W_{20}^c>2.45$. This
suggests that, when IRTF is used for measuring extragalactic distance
scale, these intrinsically fainter galaxies should be excluded from the
fitting procedure so as not to introduce unnecessary systematic error
to the inferred distance.

In spite that the photometry error is now reduced down to $<0.1$\,mag
which is negligible compared with the apparent scatter of the TF
relation, we do not obtain a significant reduction of the apparent 
scatter by using the $H_T$ instead of $H_{-0.5}^s$ (Table
\ref{tbl-3}).
This suggests that the apparent scatter has no more been dominated by
the uncertainty of the photometry, but rather by the errors of the
distance and intrinsic line width estimates, or else by the intrinsic
scatter itself. Since it is currently practically difficult to obtain a
more accurate distance for the calibrators, reducing the line width
error seems to be the most feasible and promising strategy to take to
approach the intrinsic shape of IRTF. To achieve a successful result
with this strategy, we need to obtain, among other things, definitive
numbers for the inclination of the nearby calibrators.

\subsection{Distances to UMa and Coma clusters and $H_0$}
\label{sec:distances}

Taking tentatively no account of the color problem discussed in the
previous section, we give an estimate of the cluster distances based on
the present IRTF calibrations. IRTF diagrams for the Coma cluster
galaxies is shown in Fig.\ref{fig07_coma}. Nearby calibrators are
overplotted, and the distance and the slope of IRTF for the composite 
sample are obtained in the same manner as described for UMa galaxies in 
Fig.\ref{fig07_uma}. In what follows the distances (and the slopes) are 
derived from only galaxies with $\log W_{20}^c\ge 2.45$ because of the 
reason discussed in section \ref{sec:IRTF_property}. This limitation 
takes an 
extra advantage for especially the distant Coma sample since a sample 
incompleteness bias becomes almost negligible for the brighter galaxies
(Giovanelli et al 1997b).

Estimates of the distance modulus of the UMa and Coma are summarized in
Table.\ref{tbl-4}. The quoted error is evaluated as a square root of the
sum of the squares of dispersions around the calibrated IRTF and the 
cluster IRTF each divided by the respective number of galaxies. For UMa
the distance estimates in $H_T$, $H_{-0.5}$, and $H_{19}$ show an 
excellent agreement each other, as well as with that obtained by S00 in
$I$, $m-M=31.58$\,mag. This agreement with the distance in $I$ seems
inconsistent with S00's implication that there is a severe $I-H$ offset 
between UMa galaxies and the nearby calibrators. However, our 
examination of the $I-H$ color using $H_T$ has not provided definitely
such a color offset for UMa galaxies yet (Fig.\ref{fig06}).

Estimates of the Coma cluster
distance in $H_T$ and $H_{-0.5}$ exhibit a nominal discrepancy
amounting to 0.11\,mag. Although this is within the quoted error, we
find that this discrepancy is ascribed to this particular Coma sample
for $H_{-0.5}$. This is evident from the left
panel of Fig.\ref{fig07_coma}, where it is illustrated that the majority
of the galaxies with $H_{-0.5}$ data are located underneath the 
regression line line of the IRTF in $H_T$. Indeed, if the $H_T$ sample 
is biased analogous to the $H_{-0.5}$ sample, i.e., limited to those 
with $H_{-0.5}$ measurements available, the distance modulus of Coma in 
$H_T$ is increased up to $m-M=35.08$\,mag (the fourth row in 
Table.\ref{tbl-4}), which is consistent with the biased distance 
inferred in $H_{-0.5}$. 

It is worth commenting that this tendency found for the $H_{-0.5}$ 
subsample of Coma is also observed in the TF diagram in $I_T$, i.e., 
Coma galaxies with $H_{-0.5}$ measurements available tend to be found
below the mean regression line in $I_T$. This was assured by identifying
$H_{-0.5}$-measured galaxies (Bothun et al 1985) in Giovanelli et al 
(1997a)'s sample and drawing the TF diagram in $I_T$. Similar 
identification and check were carried out for ACO\,1367, 2634, 400, 
NGC\,383 group, NGC\,507 group, Cancer, and Pegasus, but such a trend 
was not observed in these samples.

We see a shallower slope of $-5.66$ for the $H_T$ sample of UMa. This 
should be apparently ascribed to NGC\,3718 which has a largest deviation
in $H_T$ among the UMa galaxies. Peletier \& Willner (1993) found the 
similar deviation for this galaxy and argued a plausible reason, in 
terms of its optical morphological peculiarity, of justifying its 
exclusion from the TF sample.

We see in Fig.\ref{fig05} that there is a type dependence in 
$H_{-0.5}-H_T$. Although a KS test suggests no significant difference 
(at a $>70$\% confidence level) of the morphological type distribution 
between the calibrators, Coma galaxies, and UMa galaxies, we examine the
galaxy type dependence of the distance estimates. We divide the cluster 
galaxies into two subsamples of
$T\le 4$ and $T\ge 5$ and estimate the cluster distances separately 
using these subsamples and nearby calibrators divided into two as well.
The result is shown in the fifth and sixth rows in Table.\ref{tbl-4}. 
For UMa galaxies the result is consistent with that obtained from the
entire analysis. For Coma galaxies, we obtain a nominally shorter 
distance $m-M=34.70$, nevertheless it is still consistent with the
global value if we consider the large error associated to this poor
statistics.

We adopt the distance modulus $m-M=34.94$\,mag obtained in $H_T$ as our
best estimate of the Coma distance. This number, compared with
$m-M=34.74$\,mag obtained by S00 in $I$, is larger by 0.2\,mag. This
discrepancy is less than $2\sigma$ confidence level from our error
budget. However, being consistent with what is expected from the $I-H$
color problem discussed in the previous section, this discrepancy may
not be considered as an error of a random origin.

Using $m-M=34.94$\,mag of the Coma cluster and the recession velocity of
$V=7143$\,km\,s$^{-1}$ in the CMB reference frame, and assuming that the
deviation from the Hubble flow is negligible, we obtain
$H_0=73\pm 4$\,km\,s$^{-1}$\,Mpc$^{-1}$. This value is nominally
consistent with the global value of $H_0$ concluded by S00.

\section{Summary}

Infrared $J$ and $H$ surface photometry have been carried out for 12
galaxies having a Cepheid distance available via HST observations. Using
the data we have calibrated IRTFs in the $J$- and $H$-bands.
Photometric properties of $H_{-0.5}$ and colors relevant to $H$ have
been examined. The calibrated IRTFs are applied to the Coma and Ursa
Major galaxies.

The main results we have obtained are as follows;

(1) The surface photometry results are given in Table \ref{tbl-2}.
Calibrations of $J$ and $H$ IRTFs are summarized in Table \ref{tbl-3}.

(2) The offset of $H_{-0.5}$ from $H_T$ shows no significant dependence
on the galaxy axial ratio. It is galaxy-type dependent such that
$H_{-0.5}-H_T$ is larger by 0.15\,mag for late type spirals ($T>=5$)
than early ones ($T<=4$).

(3) The nearby calibrators are redder in $(I-H)_T$ than the Coma
galaxies. This is analogous to the property in $I_T-H_{-0.5}$ claimed by
S00, but such a trend is not clearly visible for UMa mostly because of
poor statistics.
It is yet unclear whether the color offset is caused alone by any
peculiarity of the $H$ luminosity.

(4) It is shown in all the magnitude systems that the IRTF 
changes its slope at $\log W_{20}^c\sim 2.45$. This phenomenon is
observed for the Ursa Major galaxies as well. The apparent scatter is
not significantly improved even if we use surface photometric magnitudes
instead of $H_{-0.5}$.

(5) From the application of the present calibration of IRTF to the Coma
cluster, we obtain its distance to be $m-M=34.94\pm 0.13$\,mag. With its
CMB-rest recession velocity we obtain 
$H_0=73\pm 4$\,km\,s$^{-1}$\,Mpc$^{-1}$. It should be noted here that 
the distance is given no account of the possible color offset mentioned 
in (4) hence can be overestimated by $\sim 0.2$\,mag. 

\acknowledgments
We are deeply indebted to Takashi Ichikawa and Kenshi Yanagisawa for
their lengthy efforts to develop and improve the large-format infrared
camera KONIC to be available on the Kiso Schmidt. Kenshi Yanagisawa has
kindly provided us with a psf image of KONIC. Thanks are due to the
staff members of the Kiso Observatory for their technical supports
during the observations. The literature data used in this study was made
available in the electronic form from the Astronomical Data Analysis
Center at National Astronomical Observatory of Japan.
We thank the referee for many valuable comments which improved the
manuscript.
\clearpage

\setlength{\tabcolsep}{1mm}

\begin{deluxetable}{clcrcccc}
\footnotesize
\tablecaption{Sample galaxies and literature data\label{tbl-1}}
\tablehead{
\colhead{NGC}					              &
\colhead{$m-M$}					              &
\colhead{Type\tablenotemark{(a)}}		      &
\colhead{$D_{25}$\tablenotemark{(a)}}		  &
\colhead{$i$\tablenotemark{(b)}}	          &
\colhead{$\log W_{20}^c$\tablenotemark{(b)}}  &
\colhead{$A_{G+i}^{(J)}$\tablenotemark{(c)}}  &
\colhead{$A_{G+i}^{(H)}$\tablenotemark{(c)}}\nl
						&
\scriptsize [mag]				&
						&
\scriptsize [arcmin]			&
						&
\scriptsize [km\,s$^{-1}$]	&
\scriptsize [mag]				&
\scriptsize [mag]
}
\startdata
925 & 29.84 (.08)\tablenotemark{(1)} & 7 & 10.47 & 56 & 2.420 & 0.22 & 0.14\nl
2541& 30.47 (.08)\tablenotemark{(2)} & 6 &  6.31 & 62 & 2.370 & 0.22 & 0.14\nl
3198& 30.80 (.06)\tablenotemark{(3)} & 5 &  8.51 & 68 & 2.531 & 0.32 & 0.21\nl
3319& 30.78 (.12)\tablenotemark{(4)} & 6 &  6.17 & 58 & 2.405 & 0.17 & 0.11\nl
3351& 30.01 (.08)\tablenotemark{(5)} & 3 &  7.41 & 45 & 2.586 & 0.14 & 0.10\nl
3368& 30.20 (.10)\tablenotemark{(6)} & 2 &  7.59 & 49 & 2.674 & 0.19 & 0.11\nl
4414& 31.41 (.10)\tablenotemark{(7)} & 5 &  3.63 & 46 & 2.743 & 0.30 & 0.19\nl
4535& 31.10 (.07)\tablenotemark{(8)} & 5 &  7.08 & 51 & 2.586 & 0.19 & 0.11\nl
4536& 30.95 (.08)\tablenotemark{(9)} & 4 &  7.59 & 69 & 2.562 & 0.36 & 0.23\nl
4639& 32.03 (.22)\tablenotemark{(10)}& 4 &  2.75 & 49\tablenotemark{(a)} & 2.619\tablenotemark{(a)}& 0.18 & 0.11\nl
4725& 30.57 (.08)\tablenotemark{(11)}& 2 & 10.72 & 62 & 2.671 & 0.15 & 0.10\nl
7331& 30.89 (.10)\tablenotemark{(12)}& 3 & 10.47 & 69 & 2.746 & 0.50 & 0.31
\enddata

\tablenotetext{}{The error of $m-M$ is given in parentheses.}
\tablenotetext{(a)}{From RC3}
\tablenotetext{(b)}
 {From Sakai et al (2000) except for
  NGC\,4639 which is taken from
  Huchtmeier \& Richter (1989) for $\log W_{20}^c$ and from
  RC3 for $i$.}
\tablenotetext{(c)}
 {From Schlegel et al (1998) for the Galactic extinction $A_G$ and Tully
  et al (1998) for the internal extinction $A_i$. See text for details.}
\tablenotetext{}{References for $m-M$ ---
(1)Silbermann et al. 1996, (2)Ferrarese et al. 1998,
(3)Kelson et al. 1999,     (4)Sakai et al. 1999,
(5)Graham et al. 1997,     (6)Tanvir et al. 1995,
(7)Turner et al. 1998,     (8)Macri et al. 1999,
(9)Saha et al. 1996,      (10)Saha et al. 1997,
(11)Gibson et al. 1999,   (12)Hughes et al. 1998}

\end{deluxetable}

\begin{deluxetable}{r@{ }r@{ }r@{ }c@{ }c@{ }c@{ }c@{ }r@{ }r@{ }r@{ }c@{ }c@{ }c@{ }c}
\footnotesize
\tablecaption{Photometry result\label{tbl-2}}
\tablehead{
\colhead{\scriptsize NGC} &
\colhead{\scriptsize $J_T$} &
\colhead{\scriptsize $J_{22}$} &
\colhead{\scriptsize $D_{J22}$} &
\colhead{\scriptsize $r_e^{(J)}$} &
\colhead{\scriptsize $R_{J22}$} &
\colhead{\scriptsize $H_T$} &
\colhead{\scriptsize $H_{-0.5}^s$} &
\colhead{\scriptsize $H_{19}$} &
\colhead{\scriptsize $H_{21}$} &
\colhead{\scriptsize $D_{H19}$} &
\colhead{\scriptsize $D_{H21}$} &
\colhead{\scriptsize $r_e^{(H)}$} &
\colhead{\scriptsize $R_{H21}$}
}
\startdata
\scriptsize  925 & \scriptsize  8.47 (.06) & \scriptsize  8.82 (.03) & \scriptsize 6.24 & \scriptsize 3.10 & \scriptsize .55 &
\scriptsize 7.54 (.05) &
\scriptsize  8.34 (.02) & \scriptsize  9.90 (.02)& \scriptsize  8.03 (.02)& \scriptsize 1.51 & \scriptsize 6.48 & \scriptsize 3.80 & \scriptsize .54 \nl
\scriptsize 2541 & \scriptsize 10.30 (.11) & \scriptsize 10.78 (.03) & \scriptsize 2.95 & \scriptsize 1.75 & \scriptsize .52 &
\scriptsize 9.85 (.11) &
\scriptsize 10.34 (.02) & \scriptsize 12.47 (.02)& \scriptsize 10.43 (.02)& \scriptsize 0.40 & \scriptsize 2.41 & \scriptsize 1.47 & \scriptsize .47 \nl
\scriptsize 3198 & \scriptsize  8.71 (.06) & \scriptsize  8.83 (.03) & \scriptsize 6.06 & \scriptsize 1.73 & \scriptsize .33 &
\scriptsize 7.93 (.05) &
\scriptsize  8.39 (.02) & \scriptsize  8.80 (.02)& \scriptsize  8.16 (.02)& \scriptsize 2.68 & \scriptsize 5.57 & \scriptsize 1.93 & \scriptsize .34 \nl
\scriptsize 3319 & \scriptsize  9.91 (.11) & \scriptsize 10.48 (.03) & \scriptsize 4.12 & \scriptsize 2.50 & \scriptsize .37 &
\scriptsize 9.15 (.11) &
\scriptsize 10.25 (.02) & \scriptsize 12.27 (.02)& \scriptsize 10.07 (.02)& \scriptsize 0.62 & \scriptsize 3.65 & \scriptsize 2.73 & \scriptsize .32 \nl
\scriptsize 3351 & \scriptsize  7.39 (.06) & \scriptsize  7.51 (.03) & \scriptsize 6.99 & \scriptsize 2.23 & \scriptsize .76 &
\scriptsize 6.90 (.05) &
\scriptsize  7.37 (.02) & \scriptsize  7.39 (.02)& \scriptsize  6.96 (.02)& \scriptsize 2.50 & \scriptsize 5.81 & \scriptsize 1.73 & \scriptsize .70 \nl
\scriptsize 3368 & \scriptsize  6.97 (.05) & \scriptsize  7.06 (.02) & \scriptsize 7.51 & \scriptsize 1.68 & \scriptsize .69 &
\scriptsize 6.24 (.05) &
\scriptsize  6.75 (.02) & \scriptsize  6.71 (.02)& \scriptsize  6.33 (.02)& \scriptsize 3.12 & \scriptsize 7.46 & \scriptsize 1.73 & \scriptsize .70 \nl
\scriptsize 4414 & \scriptsize  7.78 (.05) & \scriptsize  7.84 (.02) & \scriptsize 4.51 & \scriptsize 1.02 & \scriptsize .76 &
\scriptsize 7.14 (.05) &
\scriptsize  7.70 (.02) & \scriptsize  7.41 (.02)& \scriptsize  7.23 (.02)& \scriptsize 2.30 & \scriptsize 3.91 & \scriptsize 0.90 & \scriptsize .67 \nl
\scriptsize 4535 & \scriptsize  7.81 (.05) & \scriptsize  8.03 (.02) & \scriptsize 6.34 & \scriptsize 2.81 & \scriptsize .74 &
\scriptsize 7.31 (.06) &
\scriptsize  8.33 (.03) & \scriptsize  8.50 (.03)& \scriptsize  7.62 (.03)& \scriptsize 2.27 & \scriptsize 5.56 & \scriptsize 2.77 & \scriptsize .68 \nl
\scriptsize 4536 & \nodata          & \nodata          & \nodata    & \nodata   & \nodata   &
\scriptsize 7.52 (.05) &
\scriptsize  8.14 (.02) & \scriptsize  8.44 (.02)& \scriptsize  7.79 (.02)& \scriptsize 2.09 & \scriptsize 6.44 & \scriptsize 1.93 & \scriptsize .39 \nl
\scriptsize 4639 & \scriptsize  9.34 (.06) & \scriptsize  9.46 (.03) & \scriptsize 2.77 & \scriptsize 0.79 & \scriptsize .67 &
\scriptsize 8.56 (.05) &
\scriptsize  9.26 (.02) & \scriptsize  9.10 (.02)& \scriptsize  8.69 (.02)& \scriptsize 1.19 & \scriptsize 2.71 & \scriptsize 0.83 & \scriptsize .74 \nl
\scriptsize 4725 & \scriptsize  7.13 (.05) & \scriptsize  7.24 (.02) & \scriptsize 8.79 & \scriptsize 2.69 & \scriptsize .53 &
\scriptsize 6.21 (.05) &
\scriptsize  6.88 (.02) & \scriptsize  6.88 (.02)& \scriptsize  6.39 (.02)& \scriptsize 4.69 & \scriptsize 8.93 & \scriptsize 3.10 & \scriptsize .58 \nl
\scriptsize 7331 & \scriptsize  6.88 (.05) & \scriptsize  6.95 (.02) & \scriptsize 8.49 & \scriptsize 1.43 & \scriptsize .46 &
\scriptsize 6.15 (.05) &
\scriptsize  6.42 (.02) & \scriptsize  6.53 (.02)& \scriptsize  6.29 (.02)& \scriptsize 3.67 & \scriptsize 7.40 & \scriptsize 1.50 & \scriptsize .52
\enddata
\tablenotetext{}{See text for column definitions.}
\end{deluxetable}

\clearpage

\begin{deluxetable}{lrrrc}
\footnotesize
\tablecaption{IRTF Calibration Result\label{tbl-3}}
\tablehead{
\colhead{Mag.}	&
\colhead{$N$}		&
\colhead{Slope}	&
\colhead{Zero point}		&
\colhead{Apparent} \nl
& & & & \colhead{scatter}
}
\startdata

 $H_T$ (all)                  & 12 &  $-9.84$  (.36) & $-22.59$ (.23) & 0.38 \nl
 $H_T$ ($\log W_{20}^c>2.45$) &  9 &  $-7.54$  (.76) & $-22.95$ (.35) & 0.28 \nl

 $H_{-0.5}^s$ (all)                   & 12 & $-10.64$ (.34) & $-21.93$ (.22) & 0.36 \nl
 $H_{-0.5}^s$ ($\log W_{20}^c>2.45$)  & 9 &  $-8.37$ (.72) & $-22.29$ (.34) & 0.29 \nl

 $H_{21}$ (all)              & 12 & $-11.57$ (.34) & $-22.15$ (.22) & 0.40 \nl
 $H_{21}$ ($\log W_{20}^c>2.45$)  &  9 & $-8.14$ (.71) & $-22.70$ (.34) & 0.22 \nl

 $H_{19}$ ($\log W_{20}^c>2.45$)   &  9 & $-10.24$ (.72) & $-21.95$ (.34) & 0.15 \nl

 $J_T$ (all)                 & 11 &  $-9.67$ (.36) & $-21.93$ (.23) & 0.32 \nl
 $J_T$ ($\log W_{20}^c>2.45$)      &  8 &  $-8.48$ (.85) & $-22.13$ (.39) & 0.30 \nl

 $J_{22}$ (all)              & 11 & $-10.95$ (.34) & $-21.61$ (.23) & 0.34 \nl
 $J_{22}$ ($\log W_{20}^c>2.45$)   &  8 &  $-8.87$ (.82) & $-21.97$ (.38) & 0.28

\enddata
\tablenotetext{}{The error is given in parentheses.}
\end{deluxetable}

\begin{deluxetable}{l cccc c cccc}
\footnotesize
\tablecaption{Distance modulus to clusters\label{tbl-4}}
\tablehead{
\colhead{Mag.}	& &
\colhead{Coma}  & & & & &
\colhead{UMa}	\nl
\cline{2-5}
\cline{7-10}
 & $m-M$ & Slope & $\sigma$ & $N$ & & $m-M$ & Slope & $\sigma$ & $N$
}
\startdata
 $H_T$                    & 34.94 (.13)& $-7.24$ & 0.28    & 20      && 31.57 (.14)& $-5.66$ & 0.33    & 15 \nl
 $H_{-0.5}$               & 35.05 (.14)& $-7.67$ & 0.28    & 11      && 31.56 (.14)& $-7.44$ & 0.31    & 15 \nl
 $H_{19}$                 & \nodata    & \nodata & \nodata & \nodata && 31.56 (.14)& $-8.22$ & 0.41    & 15 \nl
 $H_T$($H_{-0.5}$ avail.) & 35.08 (.14)& $-7.53$ & 0.26    & 11      && \nodata    & \nodata & \nodata & \nodata \nl
 $H_T$($T\le 4$)           & 34.70 (.20)& $-6.08$ & 0.27    &  7      && 31.59 (.17)& $-5.71$ & 0.36    &  9 \nl
 $H_T$($T\ge 5$)           & 35.08 (.16)& $-7.87$ & 0.25    &  4      && 31.55 (.18)& $-5.43$ & 0.29    &  6
\enddata
\tablenotetext{}{The samples include only galaxies with 
$\log W_{20}^c\ge 2.45$. The Coma $H_T$ sample ($N=20$) contains only 11 galaxies to which 
$T$ is available. The error of $m-M$ is given in parentheses.}
\end{deluxetable}

\clearpage

\begin{figure} 
\caption{The luminosity profile (upper panel) and contour maps in $J$
		 (lower left) and $H$ (lower right) for the nearby calibrator
		 galaxies. The profiles denoted by triangles and squares are for
		 $J$ and $H$, respectively. Error bars represents only the 
		 isophote fitting error. R.A. and Dec in the maps are given
		 in arcmin with an arbitrary zero point. The interval of the
		 contours is 0.5\,mag\,arcsec$^{-2}$. The faintest contour
		 levels displayed are $\mu_J=22$\,mag\,arcsec$^{-2}$ and
		 $\mu_H=21$\,mag\,arcsec$^{-2}$. Some foreground stars are
		 removed, but they are not necessarily identical between $J$ and
		 $H$ images. See text for possibly disturbed isophotes of the
		 $J$ image of NGC\,925.
	 \label{fig1}}
\end{figure}

\begin{figure}
\caption{Comparison between our growth curve (solid curve) and the
		 aperture photometry data given in the literature (open
		 squares). Both $J$ and $H$ data are displayed where available.
		 A dotted line and a filled square show the template curve and
		 the resultant $H_{-0.5}$ adopted by TB95, respectively, except
		 for NGC\,4639 for which $H_{-0.5}$ is not given and the
		 template curve is displayed with arbitrary zero point.
 	 \label{fig2}}
\end{figure}

\begin{figure}
\caption{Calibration of IRTFs for $H_T$, $H_{-0.5}$, $H_{21}$,
		 $H_{19}$, $J_T$, and $J_{22}$.
 \label{fig8}}
\end{figure}

\begin{figure}
\caption{A luminosity offset $H_{-0.5}^s-H_T$ plotted against the axial
         ratio $D_{min}/D_{maj}$ for the nearby calibrators, Coma and
		 UMa galaxies. Early spirals with $T\le 4$ are shown by circles
		 while late spirals with $T\ge 5$ are shown by squares.
		 \label{fig05}}
\end{figure}

\begin{figure}
\caption{A color $(I-H)_T$ plotted against $\log W_{20}^c$ for the
		 nearby calibrators (filled squares), Coma galaxies (stars) and
		 Ursa Major galaxies (circles). \label{fig06}}
\end{figure}

\begin{figure}
\caption{The same as Fig.\ref{fig06} but for $(B-H)_T$.  \label{fig08}}
\end{figure}

\begin{figure}
\caption{$H$-band IRTF diagrams for the Ursa Major galaxies (small 
		 squares) overplotted by the nearby calibrators (large squares).
		 A small open square represents NGC\,3718 (see section 
		 \ref{sec:distances} for detail).
		 The calibrators are vertically scaled according to the best
		 estimates of the cluster distance modulus. The $H$ magnitudes
		 used are, from the left panel to right, $H_T$, $H_{-0.5}$, and
		 $H_{19}$. 
		 \label{fig07_uma}}
\end{figure}

\begin{figure}
\caption{The same as Fig.\ref{fig07_uma} but for the Coma galaxies. The
		 $H$ magnitudes used are $H_T$ ( the left and middle panels )
		 and $H_{-0.5}$ ( the right panel ). In the left panel the
		 distance is estimated using all the Coma galaxies including
		 those without $H_{-0.5}$ data ( crosses ). In the middle panel,
		 on the other hand, the distance is estimated just using those
		 with $H_{-0.5}$ data. \label{fig07_coma}}
\end{figure}

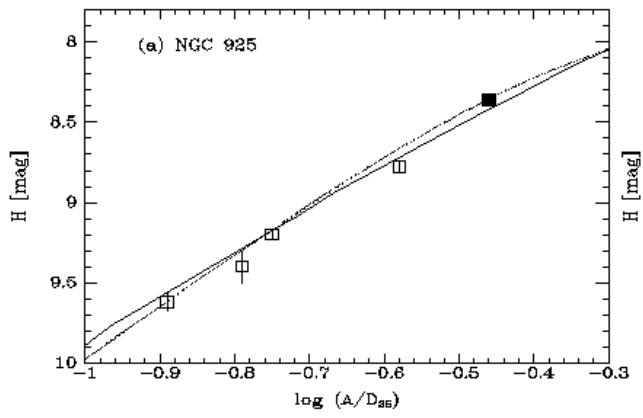
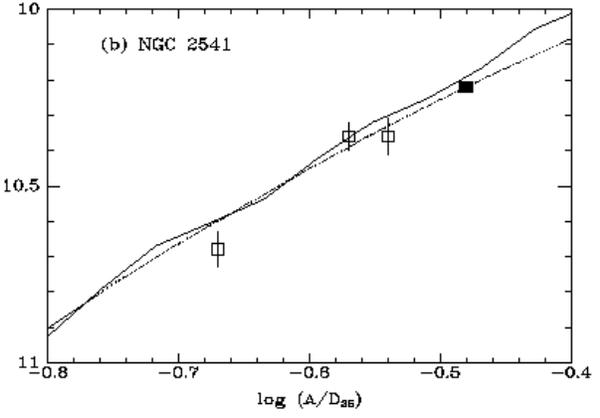
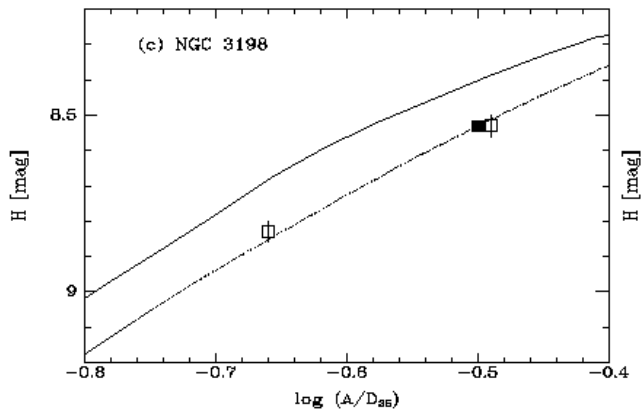
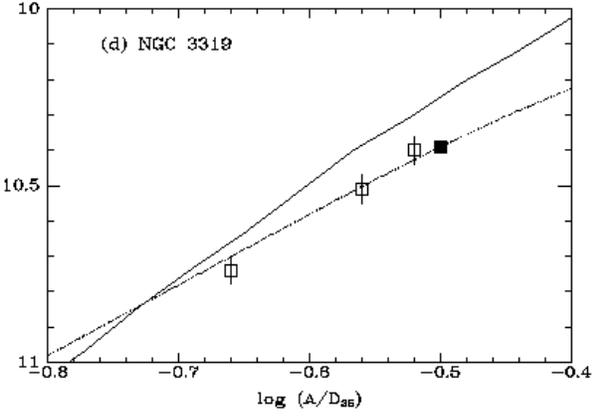
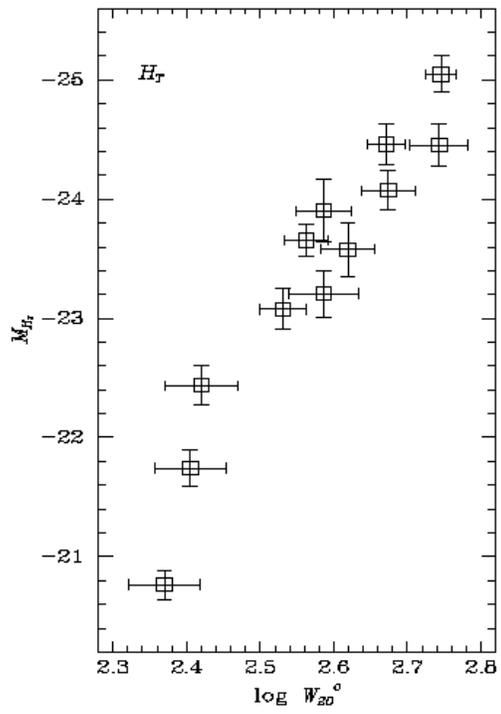
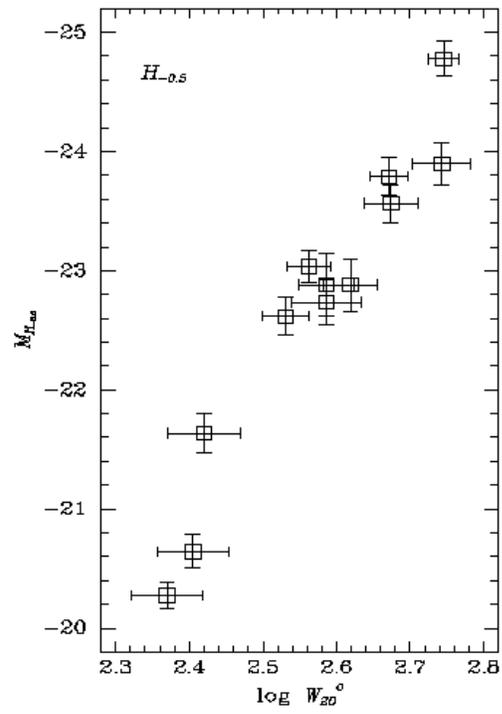
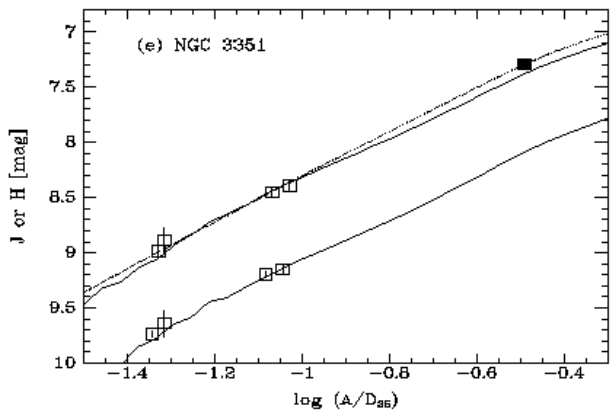
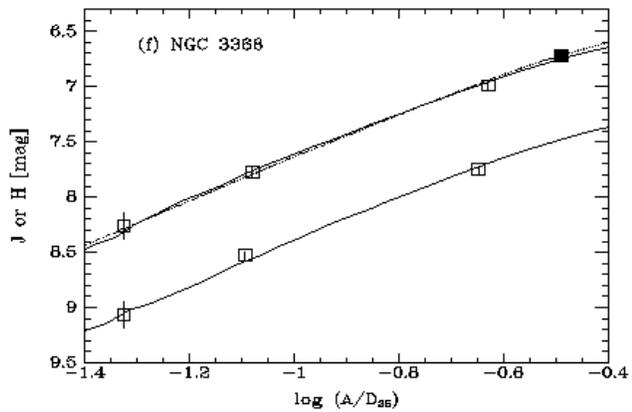
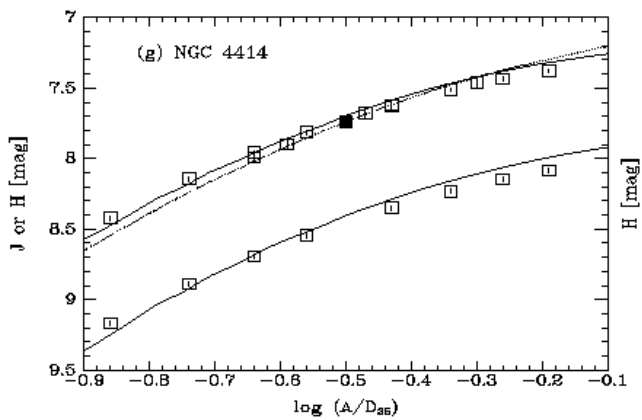
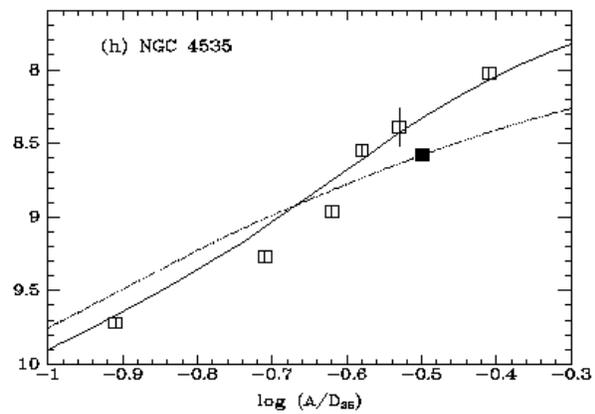
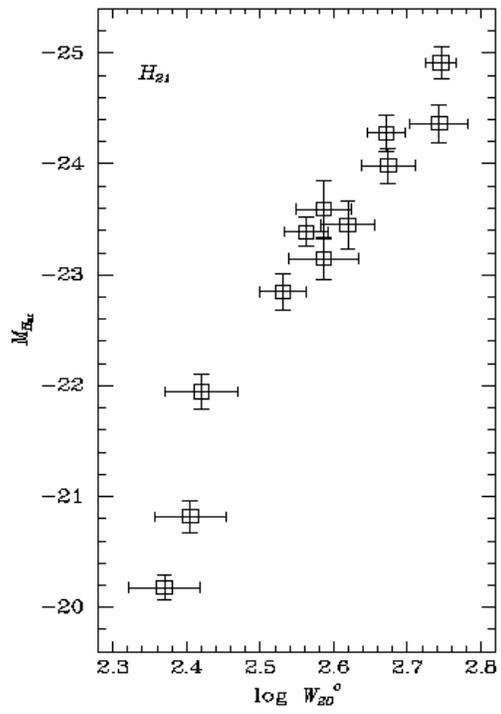
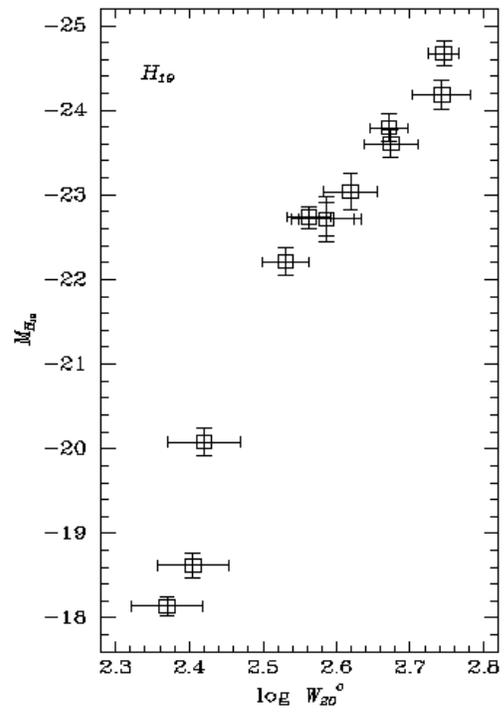
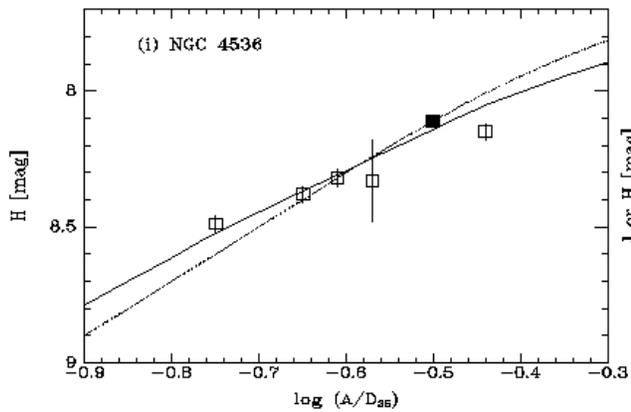
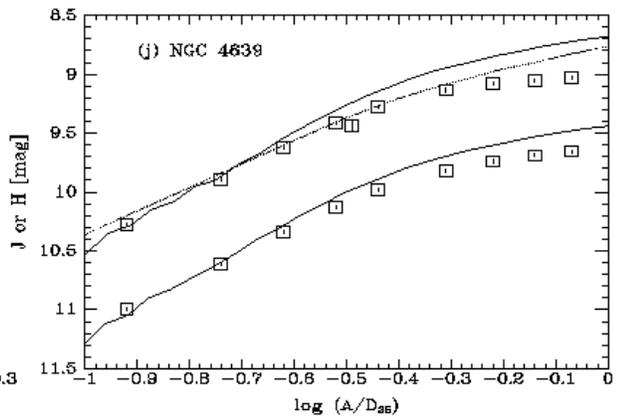
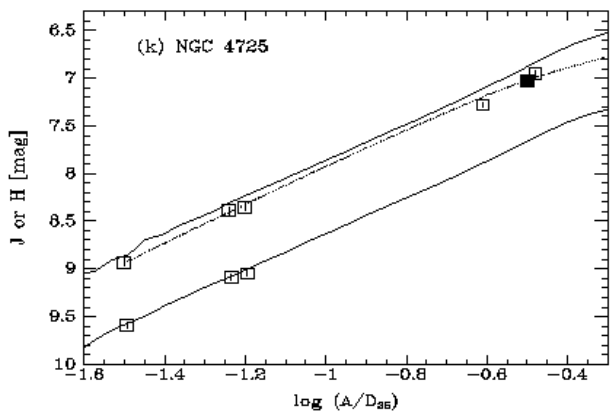
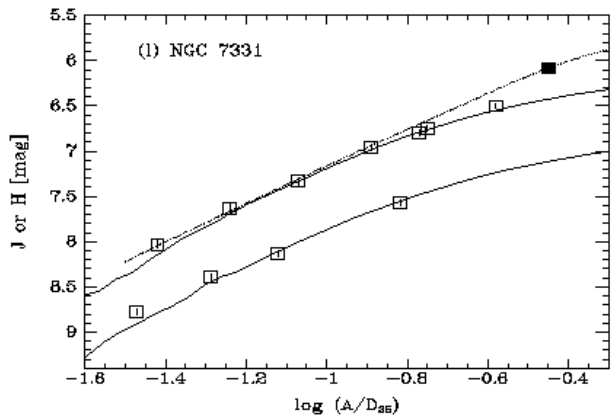
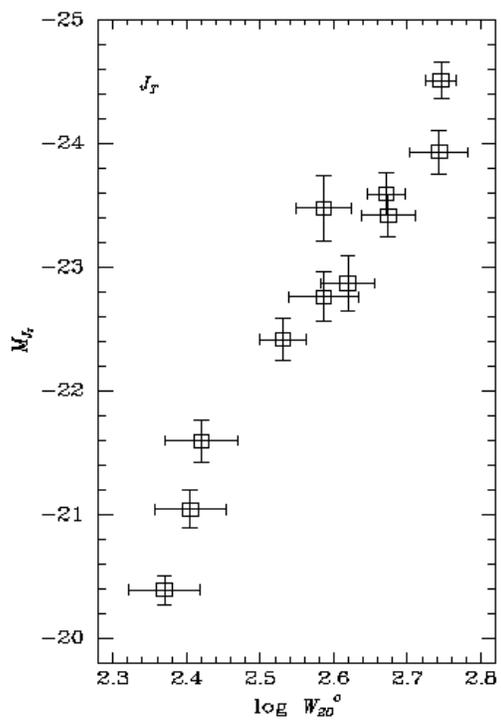
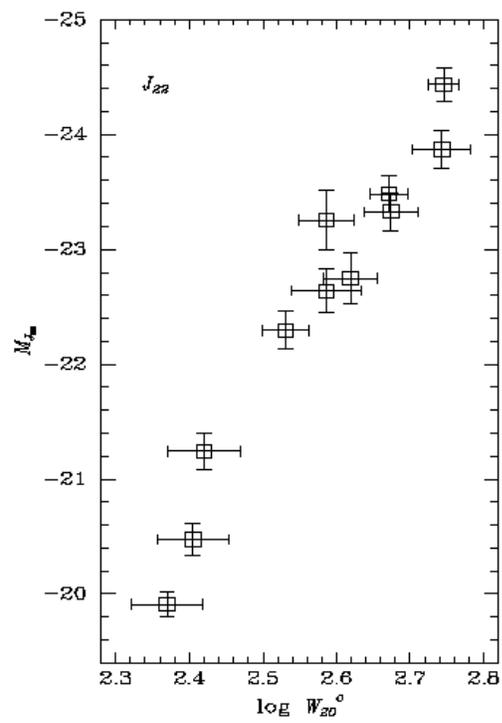
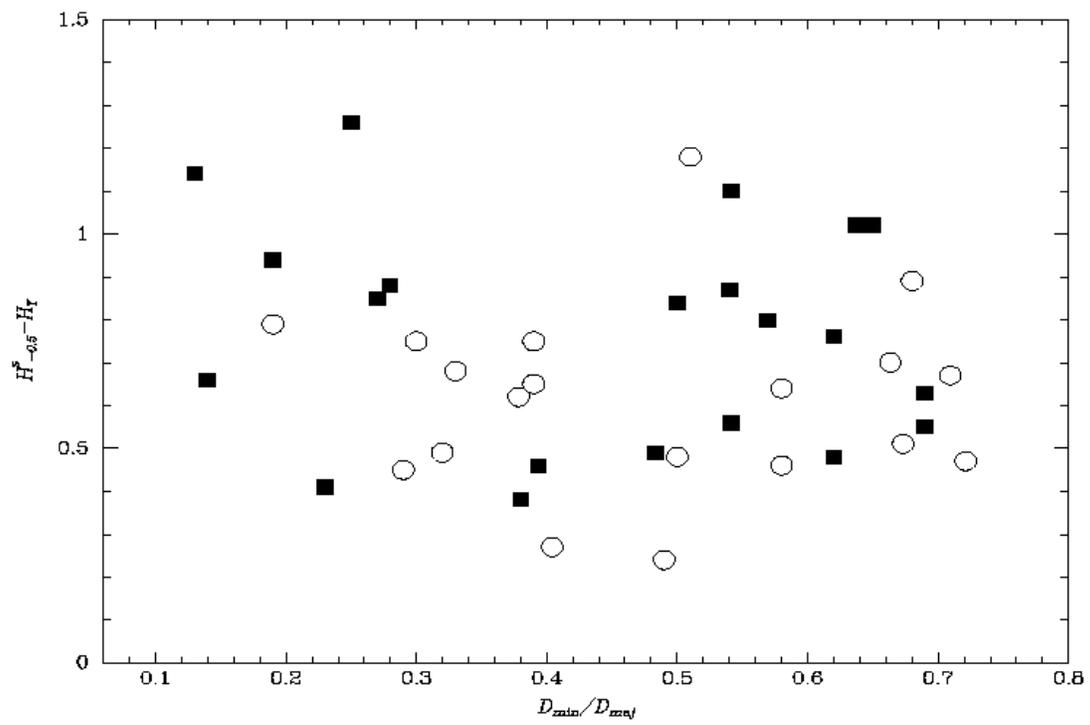
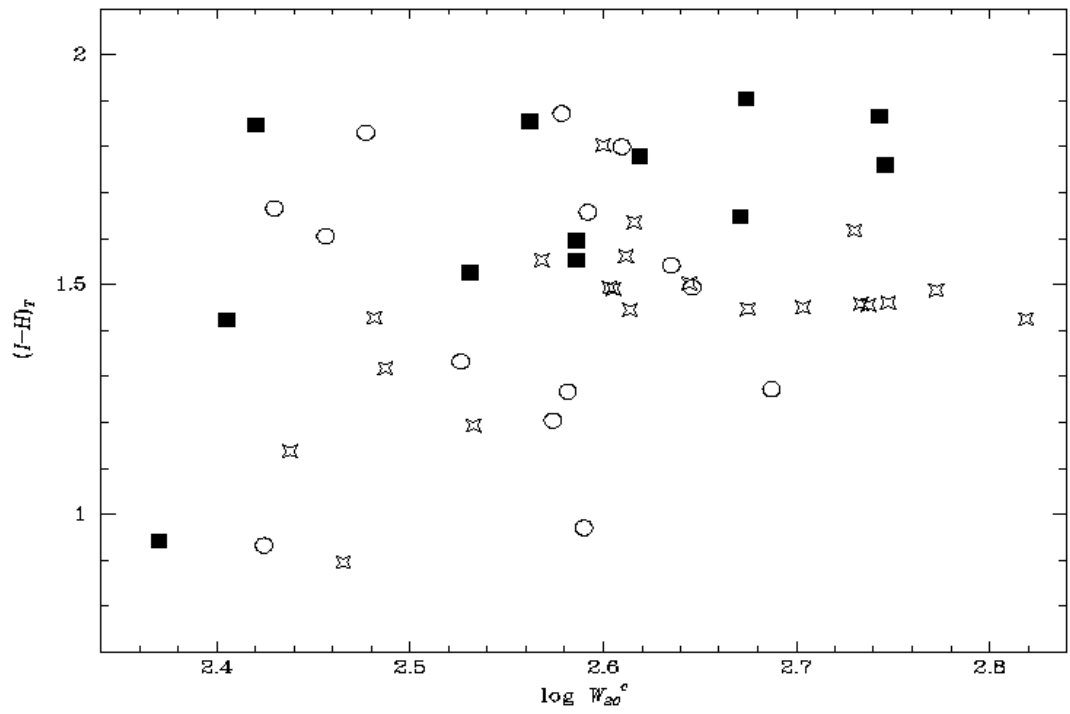
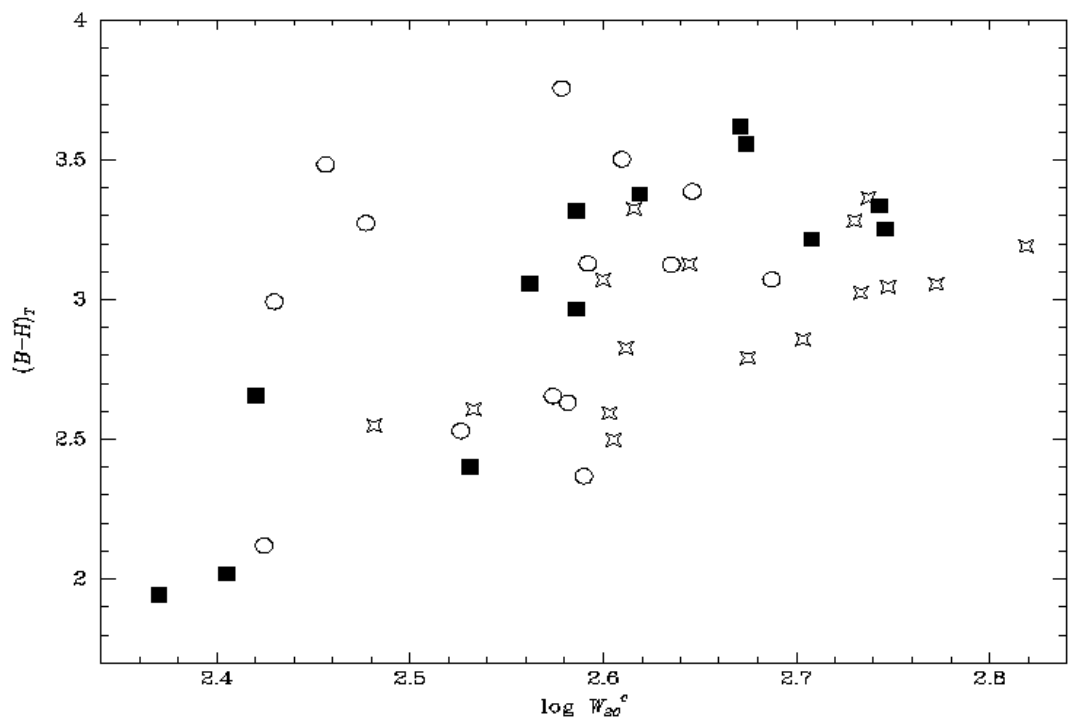
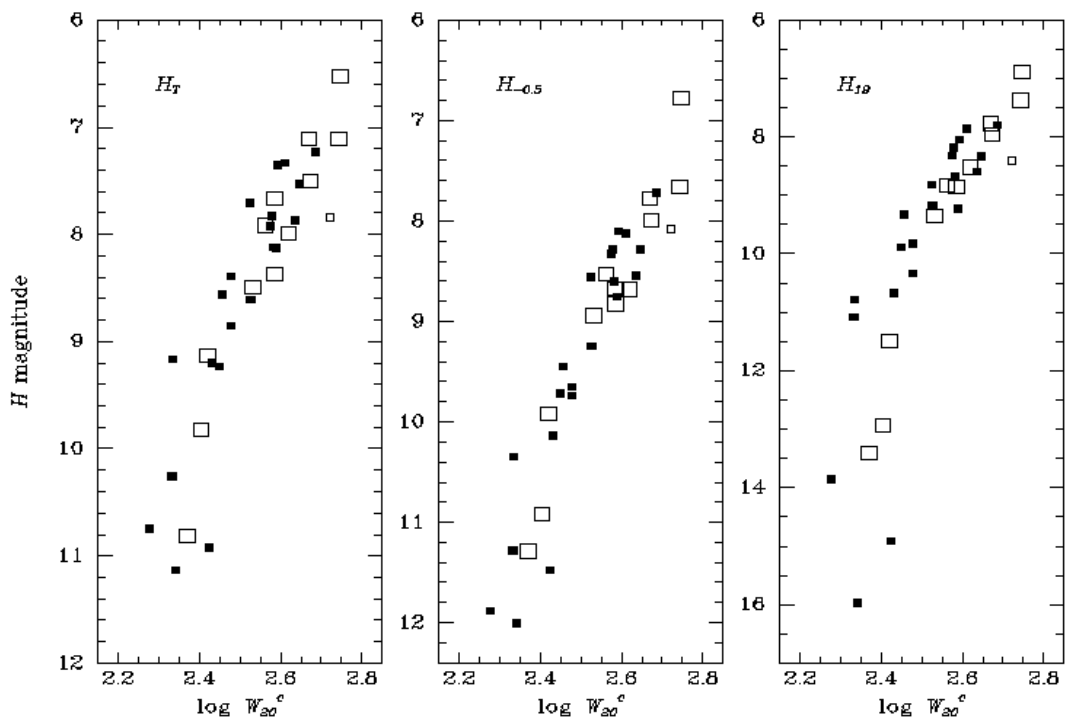
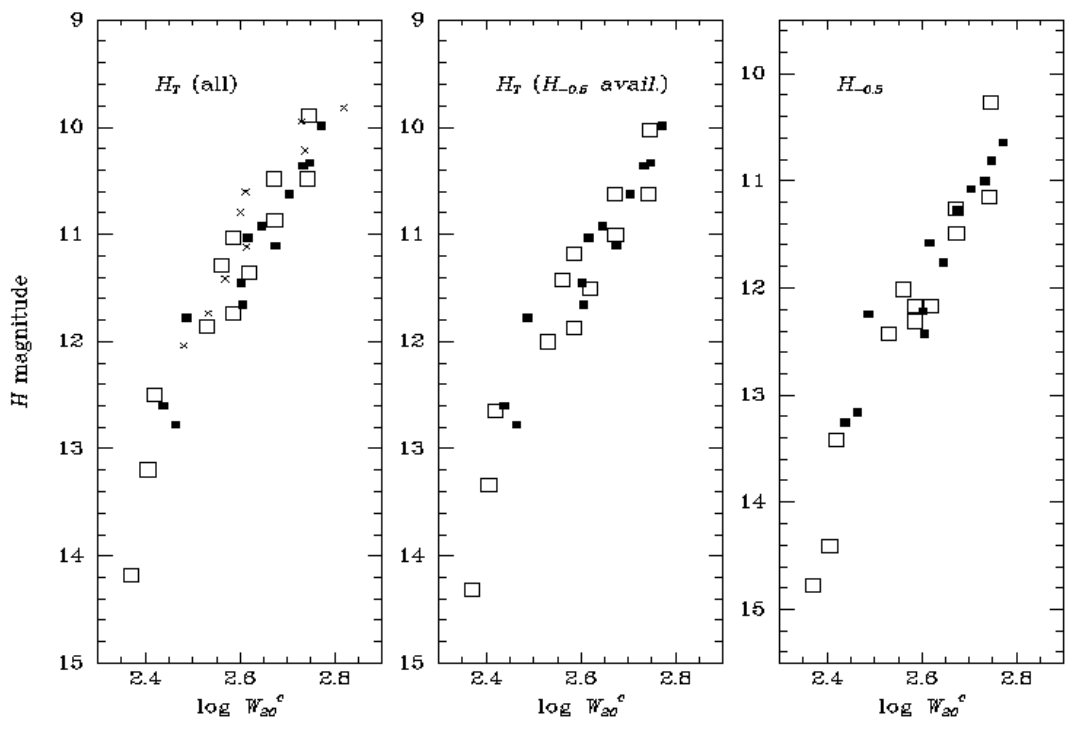
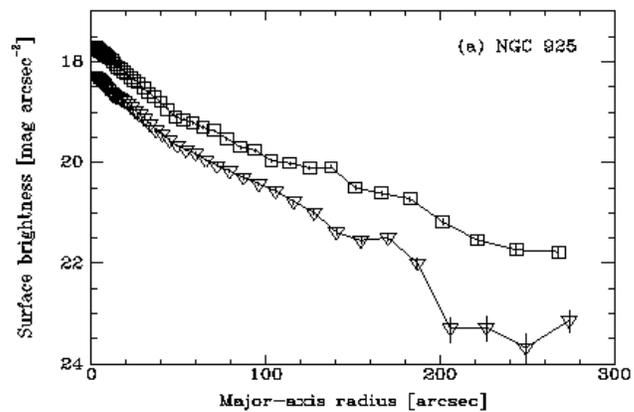
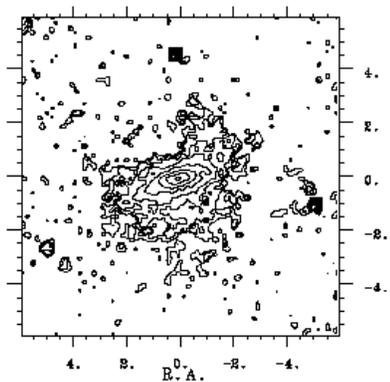
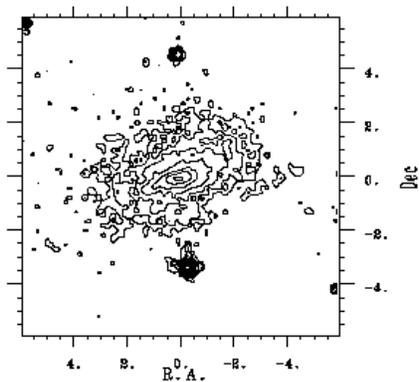
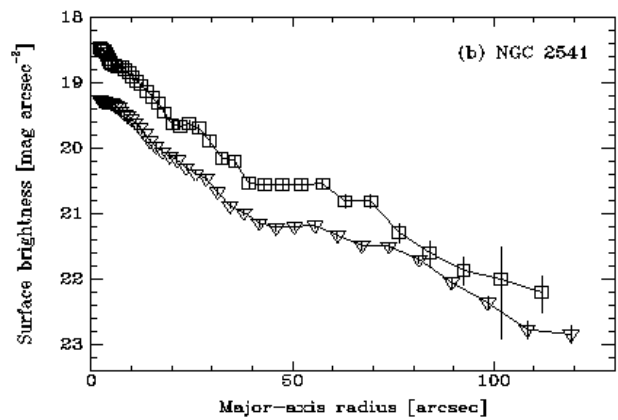
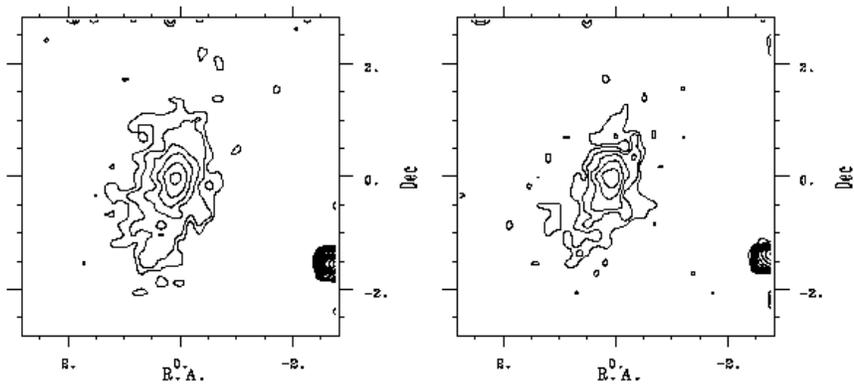
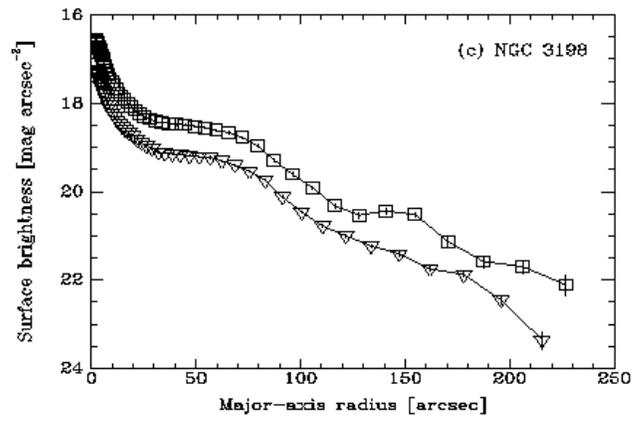
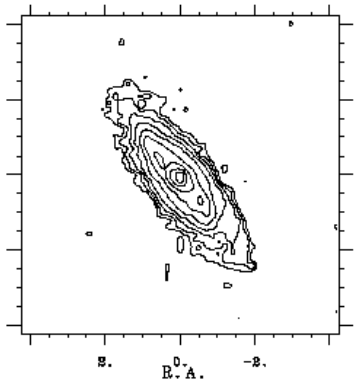
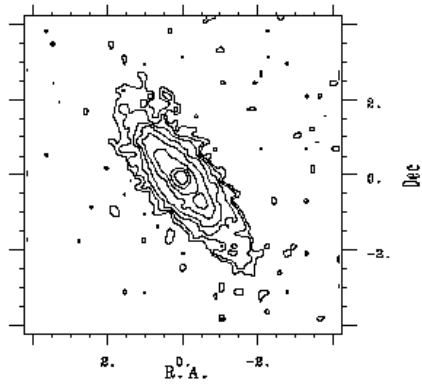
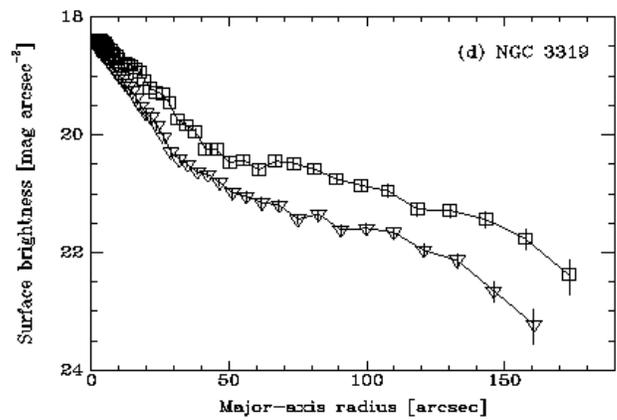
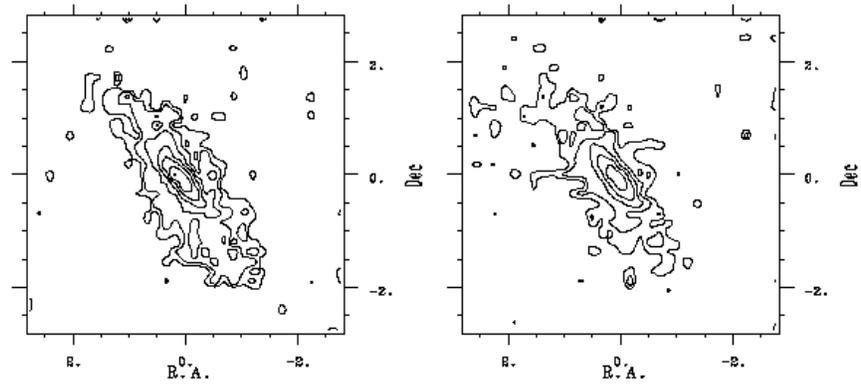
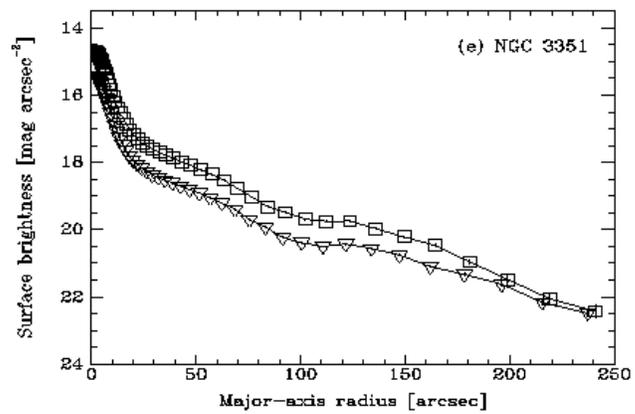
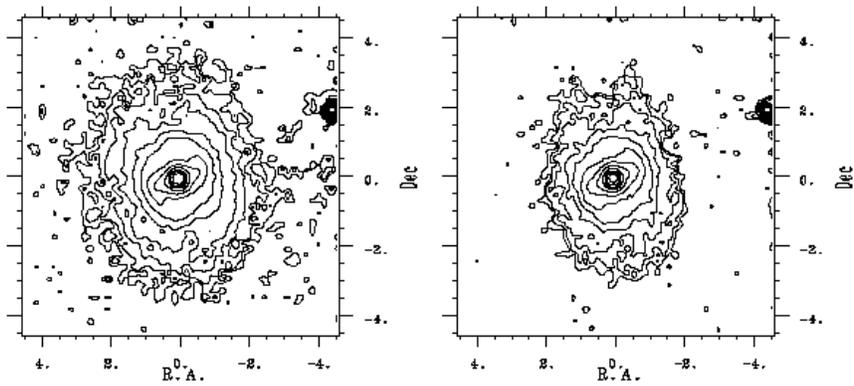
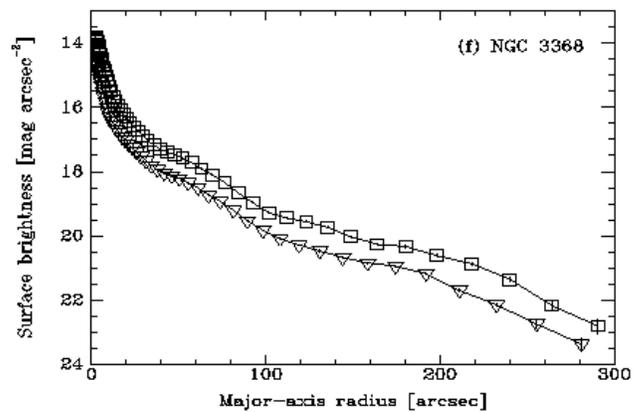
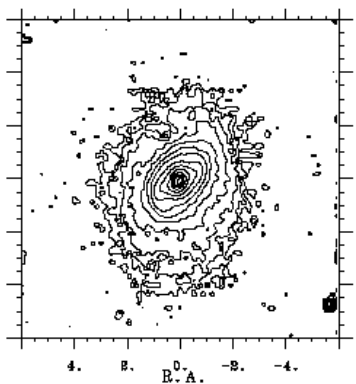
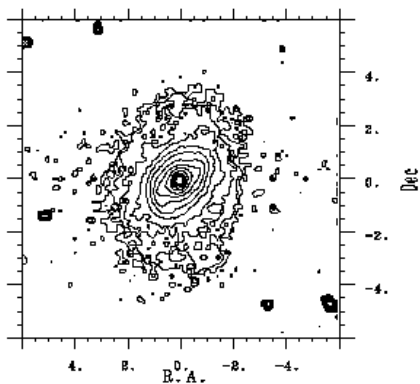
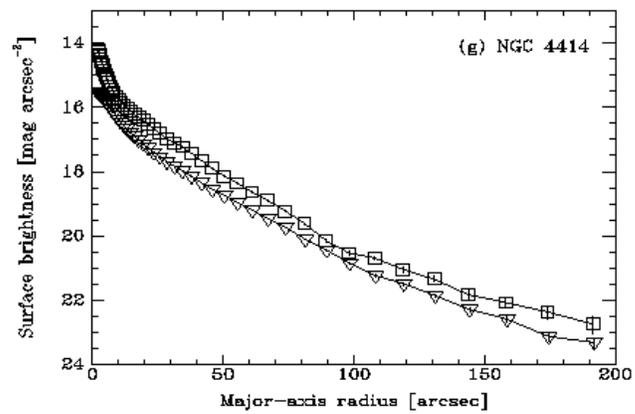
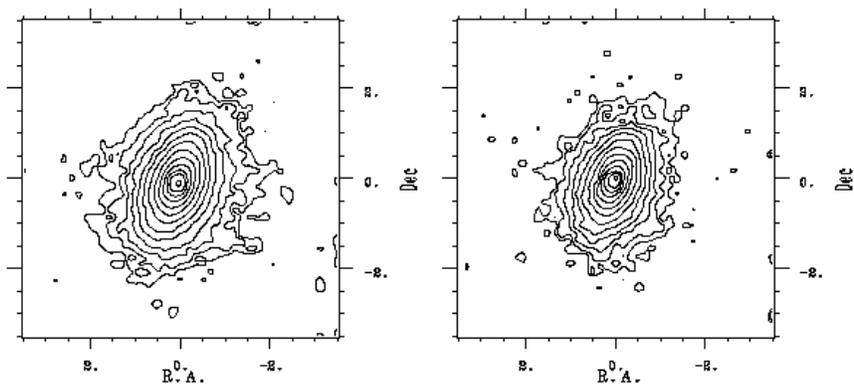
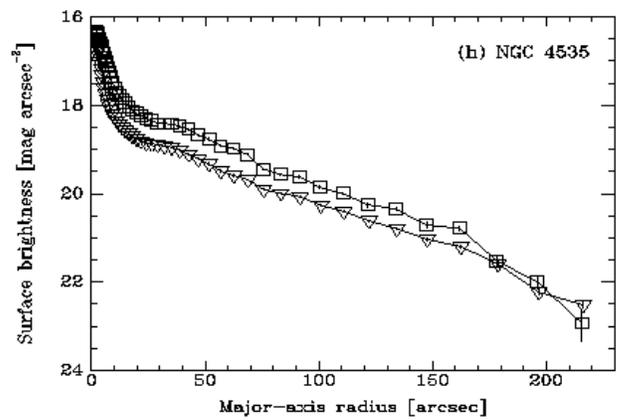
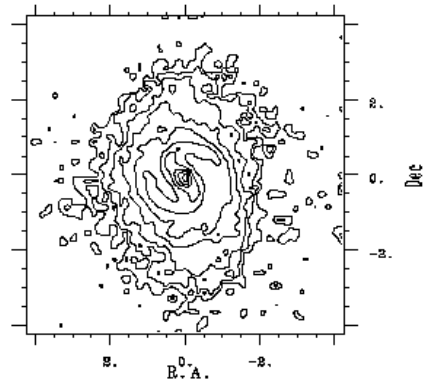
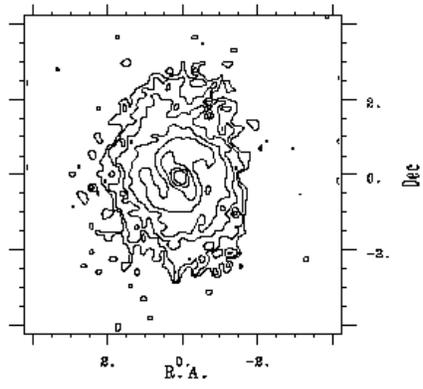
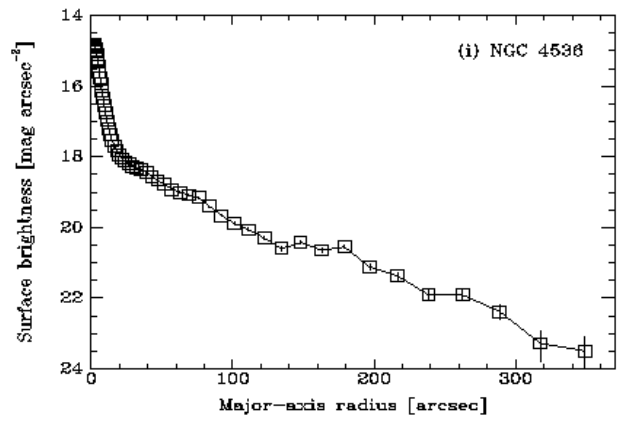
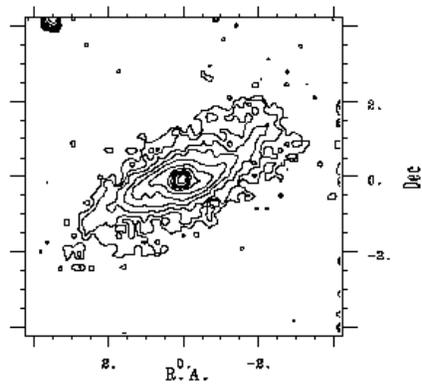
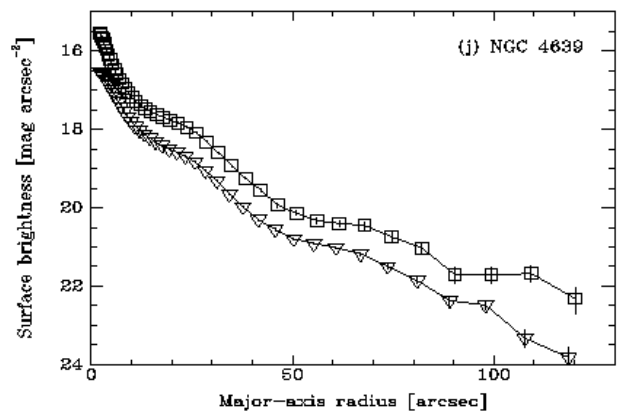
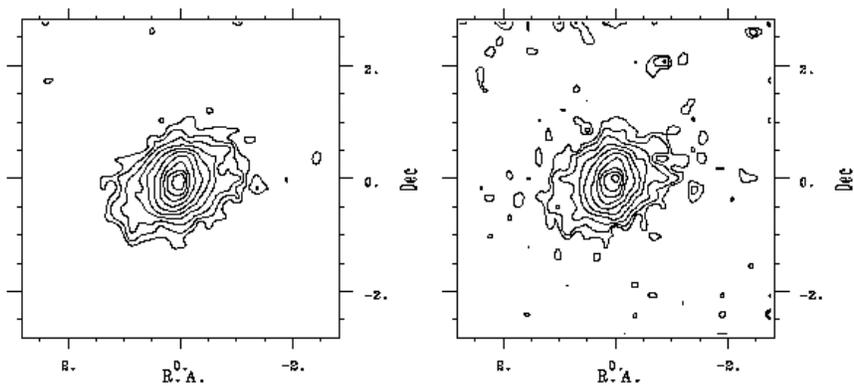
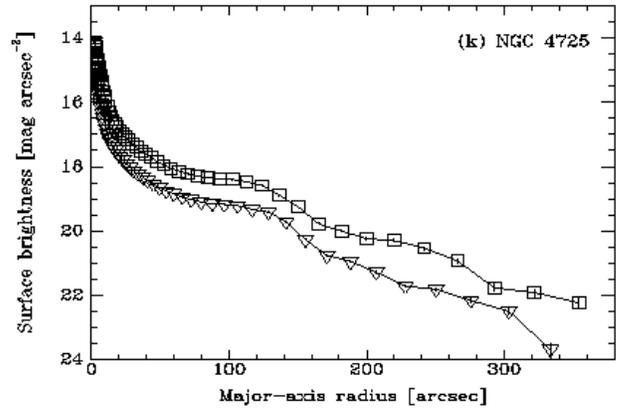
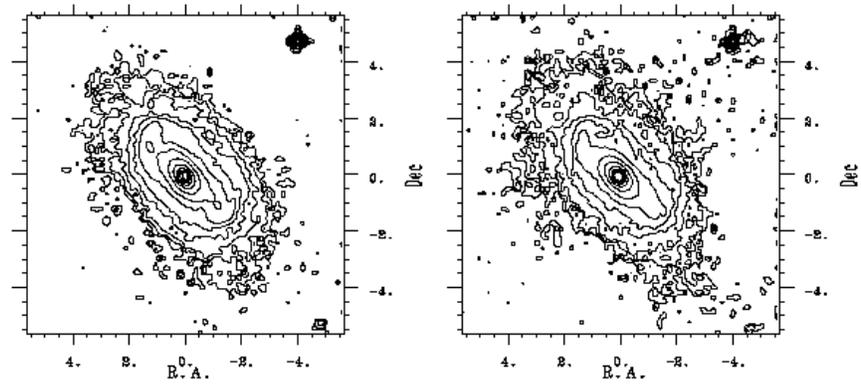
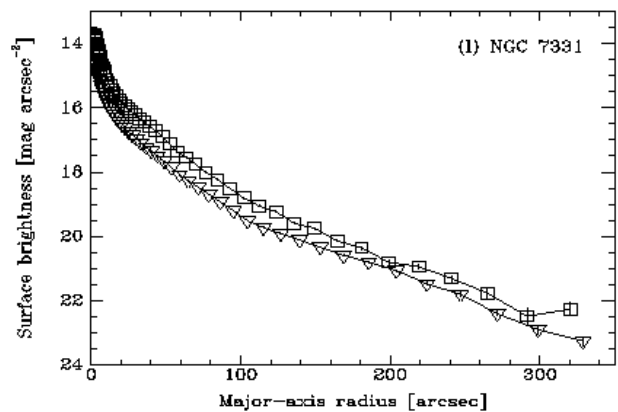
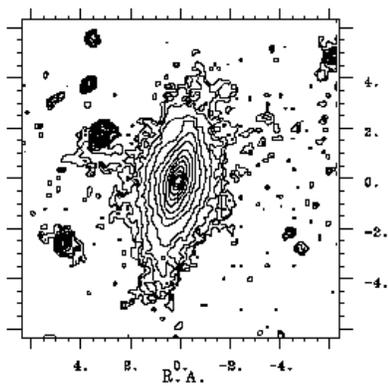
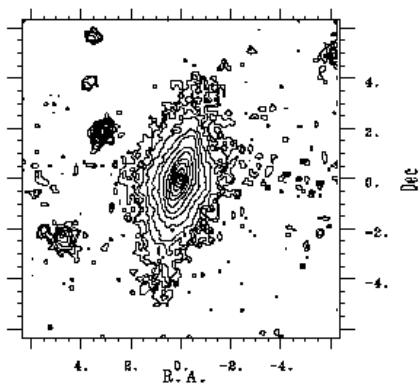
\end{document}